\documentclass[aps,prb,superscriptaddress, 12pt,amsmath,amssymb,longbibliography, preprint]{revtex4-2}

\usepackage[T1]{fontenc}

\usepackage{graphicx}
\usepackage{siunitx}	
\usepackage{amsmath,amssymb}
\usepackage{datetime}
\usepackage{booktabs}
\usepackage{multirow}
\usepackage{makecell}
\usepackage{hyperref}
\usepackage{tabularx}
\usepackage[noabbrev, capitalise, nameinlink]{cleveref}
\renewcommand{\autoref}[1]{\cref{#1}}
\usepackage{xcolor}
\hypersetup{
    colorlinks,
    linkcolor={red!50!black},
    citecolor={blue!50!black},
    urlcolor={blue!80!black}
}

\usepackage{todonotes}
\usepackage{ulem}
\usepackage{chemformula} 
\usepackage{makecell}

\begin{document}


\preprint{ver 0.1 \today\ \currenttime}

\title{Synthesis and stability of high--$T_c$ LaH$_{10\pm\delta}$ films at high pressures}

\author{Sam Cross}
\email{smc270@bath.ac.uk}
\affiliation{Department of Physics, University of Bath, Bath, BA2 7AY, UK}
\affiliation{HH Wills Laboratory, University of Bristol, Bristol, BS8 1TL, UK}

\author{William Thomas}
\affiliation{HH Wills Laboratory, University of Bristol, Bristol, BS8 1TL, UK}

\author{Lawrence Nobbs}
\affiliation{HH Wills Laboratory, University of Bristol, Bristol, BS8 1TL, UK}

\author{Rebecca Nicholls}
\affiliation{HH Wills Laboratory, University of Bristol, Bristol, BS8 1TL, UK}

\author{Oliver Lord}
\affiliation{School of Earth Sciences, University of Bristol, Bristol BS8 1RJ, UK}

\author{Qian Zhang}
\affiliation{Centre for Science at Extreme Conditions and School of Physics and Astronomy, University of Edinburgh, Edinburgh, EH9 3FD, UK}

\author{Dominique Laniel}
\affiliation{Centre for Science at Extreme Conditions and School of Physics and Astronomy, University of Edinburgh, Edinburgh, EH9 3FD, UK}

\author{Max Gerin}
\affiliation{European Synchrotron Radiation Facility (ESRF), 71 avenue des Martyrs, 38000 Grenoble, France}

\author{Bjorn Wehinger}
\affiliation{European Synchrotron Radiation Facility (ESRF), 71 avenue des Martyrs, 38000 Grenoble, France}

\author{Mohamed Mezouar}
\affiliation{European Synchrotron Radiation Facility (ESRF), 71 avenue des Martyrs, 38000 Grenoble, France}

\author{Xiaojiao Liu}
\affiliation{Diamond Light Source, Chilton, Oxfordshire, OX11 0DE, UK}

\author{Egor Koemets}
\affiliation{Diamond Light Source, Chilton, Oxfordshire, OX11 0DE, UK}

\author{Annette Kleppe}
\affiliation{Diamond Light Source, Chilton, Oxfordshire, OX11 0DE, UK}

\author{Sven Friedemann}
\affiliation{HH Wills Laboratory, University of Bristol, Bristol, BS8 1TL, UK}

\author{Jonathan Buhot}
\email{jypb20@bath.ac.uk}
\affiliation{Department of Physics, University of Bath, Bath, BA2 7AY, UK}
\affiliation{HH Wills Laboratory, University of Bristol, Bristol, BS8 1TL, UK}

\date{\today}

\begin{abstract}

High-pressure hydrides hold the record for the highest superconducting critical temperatures across all classes of superconductors. Currently lanthanum decahydride, LaH$_{10}$, exhibits the highest critical temperature among binaries, with $T_c \approx \SI{250}{\kelvin}$ at pressures between \SI{140}{}--\SI{180}{\GPa}. Here, we report the synthesis of LaH$_{10\pm\delta}$ films in two diamond anvil cells (DACs) at pressures of \SI{168}{\GPa} and \SI{176}{\GPa} via in situ laser heating of elemental lanthanum films with ammonia borane (NH$_3$BH$_3$) as the hydrogen donor. The high-symmetry \textit{fcc} lanthanum sublattice (space group $Fm\bar3m$) is resolved using synchrotron X-ray diffraction, with unit cell parameters in excellent agreement with previous studies on bulk samples. We provide confirmation of high-$T_c$ superconductivity in LaH$_{10\pm\delta}$ with highest $T_c$ of \SI{247}{\kelvin} at \SI{176}{\GPa} evidenced in electrical measurements. The characteristic suppression of superconductivity is observed in magnetic fields. Furthermore, combined diffraction and electrical measurements reveal remarkable temporal stability of both the crystal structure and the high-$T_c$ superconductivity over the full measurement period of about 300 days post laser heating. Our work establishes film precursors using physical vapour deposition (PVD) techniques as a practical route to hydride formation, opening a pathway toward the controlled synthesis of promising ternary hydrides and the integration of micro-fabricated device geometries in diamond anvil cells. 

\end{abstract}

\maketitle

\section{Introduction}

Since the discovery of high-temperature superconductivity in sulfur hydride, H$_3$S, in 2015 \cite{drozdov_conventional_2015}, numerous compressed metal hydrides have been found to exhibit very high superconducting critical temperatures, $T_c$ \cite{drozdov_conventional_2015, drozdov_superconductivity_2019, somayazulu_evidence_2019, osmond_clean-limit_2022, cross_high-temperature_2024, kong_superconductivity_2021, chen_high-temperature_2021, guo_unusual_2024, sun_high-temperature_2021}. The current record $T_c$, among binaries, is held by lanthanum decahydride, LaH$_{10}$, with $T_c \approx \SI{250}{\kelvin}$ at pressures above \SI{140}{\giga\pascal} \cite{drozdov_superconductivity_2019, somayazulu_evidence_2019}. In LaH$_{10}$, lanthanum atoms form an \textit{fcc} lattice and are surrounded by clathrate hydrogen networks that are stabilised by the charge transfer from electropositive atoms to promote the dissociation of molecular H$_2$ units and the formation of hydrogen cages \cite{flores-livas_perspective_2020, peng_hydrogen_2017}. This yields a high hydrogen-derived density of states at the Fermi level coupled to high-frequency phonon modes, crucial for realising high-$T_c$ superconductivity in LaH$_{10}$ and many other hydride systems \cite{drozdov_superconductivity_2019, somayazulu_evidence_2019, sun_high-temperature_2021, cross_high-temperature_2024, guo_unusual_2024, osmond_hydrogen_2026, kong_superconductivity_2021, chen_high-temperature_2021}.

Questions nevertheless remain regarding the long-term stability of $Fm\bar3m$-LaH$_{10}$ under sustained compression.
Initially, first-principles calculations within the harmonic approximation predicted the $Fm\bar3m$ clathrate phase to be dynamically stable above \SI{210}{\giga\pascal} \cite{liu_potential_2017}. Later, quantum anharmonic effects were shown to dynamically stabilise $Fm\bar3m$-LaH$_{10}$ down to \SI{129}{\giga\pascal} \cite{errea_quantum_2020}. This was consistent with experimental observations of the $Fm\bar3m$-LaH$_{10}$ phase persisting to \SI{138}{\giga\pascal}, below which a distortion to a lower-symmetry $C2/m$ structure occurs, accompanied by a rapid reduction in $T_c$ upon further decompression \cite{sun_high-temperature_2021}.
Recent nuclear magnetic resonance (NMR) measurements have raised questions concerning the long-term stability of LaH$_{10}$, reporting a highly diffusive state of the hydrogen sublattice at room temperature which was found to drive progressive dehydrogenation of the sample toward lanthanum trihydride, LaH$_3$, over the course of about 70 days \cite{zhou_diffusion-driven_2025}. It should be noted that the NMR study did not conduct X-ray diffraction (XRD) measurements to confirm the presence of $Fm\bar3m$-LaH$_{10}$. In contrast, diffraction and transport studies of bulk LaH$_{10}$ samples have shown stability of the crystal structure and superconducting critical temperature of LaH$_{10}$ over timescales exceeding four years \cite{minkov_long-term_2026}. Further investigations combining structural and transport probes are therefore desirable to resolve this controversy.

The synthesis and characterisation of hydrides are challenging owing to the extreme pressures required. Typically hydrides are synthesised via in situ laser heating of bulk elemental precursors with a hydrogen donor compound, such as pure H$_2$, ammonia borane or paraffin oil, at high pressures inside a diamond anvil cell (DAC). While many experimental successes have been reported, the synthesis of clean single-phase hydride samples remains challenging owing to multiple competing phases that may coexist at similar pressures. Phase coexistence in the lanthanum hydride system has been evidenced in experimental studies \cite{laniel_high-pressure_2022} and corroborated by calculations that identify competing thermodynamically stable phases \cite{peng_hydrogen_2017, liu_potential_2017}. Partial or inhomogeneous hydrogenation during laser heating coupled with pressure and thermal gradients can further promote the formation of multiple phases within the same sample. Nevertheless, targeting the synthesis of single-phase hydrides is important for accurate characterisation of intrinsic structural and superconducting properties, and highlights a pertinent challenge for the field. 

One promising avenue toward this goal is the use of film precursors deposited directly onto the diamond anvil. Such methods have allowed successful synthesis of superconducting La$_4$H$_{23}$ films at \SI{100}{\GPa} by laser heating a lanthanum film with ammonia borane as the hydrogen donor \cite{cross_high-temperature_2024}. Film electrodes have been used to study the superconductivity in bulk H$_3$S \cite{osmond_clean-limit_2022} and in elemental yttrium films at high-pressures \cite{buhot_experimental_2020}. The use of films in hydride synthesis ensures a high ratio of hydrogen to metal, thus promoting the formation of hydrides with a high hydrogen stoichiometry. Furthermore, deposition directly onto the diamond culet simplifies sample preparation, ensures reliable contact to pre-patterned electrodes on compression, and enables better control over the initial sample geometry. While certain challenges remain, such as effective laser heating of a thin sample with strong thermal link to the anvil, and weaker XRD intensities relative to bulk samples, film techniques have greatly improved the reproducibility and success rate of our DAC preparation and synthesis attempts. Based on our experience, we estimate that the success rate in maintaining electrical contact to film samples during DAC loading and pressurising exceeds 90\%, in contrast to the lower success rates typically achieved using traditional methods, namely positioning bulk samples onto hand-cut electrodes prepared from a metallic foil \cite{balakirev2024evidence}.

Here, we report the first synthesis of high-$T_c$ $Fm\bar{3}m$-LaH$_{10\pm\delta}$ films at pressures above \SI{160}{\GPa}, formed by laser heating elemental lanthanum films with purified ammonia borane as the hydrogen donor. The target $Fm\bar{3}m$ phase was successfully obtained in two DACs prepared for this study. In section \ref{Section: LaH10 films: Structural and electronic properties} we present a structural and electrical characterisation of the synthesised films, and provide strong evidence for superconductivity in $Fm\bar{3}m$-LaH$_{10\pm\delta}$ with detailed analysis of the zero-resistance state. In section \ref{sec:Temporal stability} we present a stability study of our films by combining high-resolution X-ray diffraction and electrical transport measurements, addressing key questions regarding the long-term stability of LaH$_{10\pm\delta}$.  

\section{LaH10 films: Structural and electronic properties}
\label{Section: LaH10 films: Structural and electronic properties}

\subsection{Structural characterisation}

\begin{figure}[ht!]
    \centering
    \includegraphics[width=0.8\textwidth]{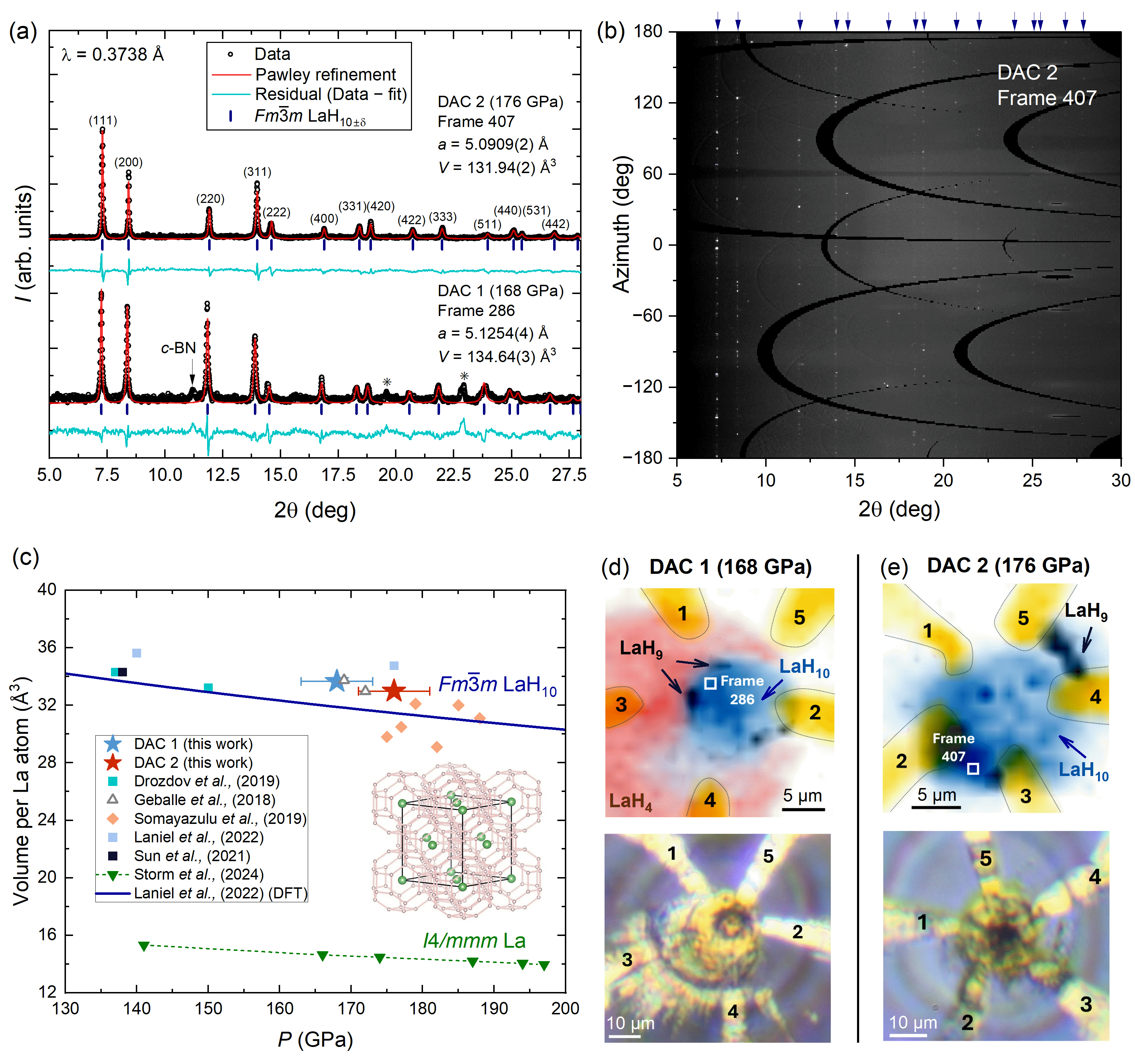}
    \caption{\footnotesize Synchrotron X-ray diffraction measurements, measured 195 days and 103 days after laser heating of DACs 1 and 2 respectively. (a) Integrated diffraction patterns demonstrating the presence of $Fm\bar3m$-LaH$_{10\pm\delta}$ in both DACs. Data are shown as open black circles, and red solid lines are Pawley refinements. Intensities have been normalised to the (111) reflection intensity for comparison. Reflections that could not be indexed are indicated with an asterisk ($\ast$). (b) Azimuthal projection of the detector image in frame 407 in DAC 2, corresponding to the integrated pattern in (a). Blue arrows indicate the $Fm\bar3m$ reflections. (c) Volume per lanthanum atom as a function of pressure, including data for LaH$_{10\pm\delta}$ samples synthesised in this work together with those in Refs. \cite{drozdov_superconductivity_2019, geballe_synthesis_2018, somayazulu_evidence_2019, laniel_high-pressure_2022, sun_high-temperature_2021}. The DFT equation of state from Ref. \cite{laniel_high-pressure_2022} is shown as the solid blue line. Experimental data for $I4/mmm$-La from Ref. \cite{storm_crystal_2024} are shown as the green inverted triangles. (d) and (e) Spatial XRD mappings and photomicrographs of both samples. White open squares on the spatial maps correspond to the positions of the frames of the respective integrated patterns in (a). Blue regions on the spatial maps correspond to intensity distribution of the $Fm\bar3m$-LaH$_{10\pm\delta}$ (311) reflection, and the red and black regions correspond to the (101) reflections of $I4/mmm$-LaH$_{4\pm\delta}$ and $P6_3/mmc$-LaH$_{9\pm\delta}$, respectively. The electrode maps are shown in yellow, corresponding to the distribution of the $Fm\bar3m$-Au (111) reflection intensity.
    }
    \label{fig:XRD_summary}
\end{figure}

Two DACs were prepared with elemental lanthanum films and ammonia borane at high-pressures as detailed in Section \ref{Experimental details}. The successful synthesis of the high-symmetry $Fm\bar3m$-LaH$_{10\pm\delta}$ phase in both DACs is evidenced using synchrotron X-ray diffraction measurements presented in Figure \ref{fig:XRD_summary}(a). Unit cell parameters were extracted for the $Fm\bar3m$ phase from Pawley refinements \cite{pawley_unit-cell_1981} and were found to be $a=\SI{5.1254(4)}{\angstrom}$ ($V=\SI{134.64(3)}{\angstrom^3}$) in DAC 1 at \SI{168}{\GPa} and $a=\SI{5.0909(2)}{\angstrom}$ ($V=\SI{131.94(2)}{\angstrom^3}$) in DAC 2 at \SI{176}{\GPa}. The azimuthal projection of the detector image corresponding to the frame employed for the Pawley refinement of DAC 2 is shown in Figure \ref{fig:XRD_summary}(b). For our polycrystalline films, typically spotty diffraction patterns were observed as in Figure \ref{fig:XRD_summary}(b) similar to those observed in previous studies of laser-heated LaH$_{10}$ samples \cite{drozdov_superconductivity_2019, sun_high-temperature_2021}. The beam diameter was \SI{0.7}{\micro\metre} in full width at half maximum (FWHM), and hence only a small number of crystallites were illuminated, and relative peak intensities for the $Fm\bar3m$ phase often deviated from those expected from a powder average. Hence, the Pawley refinement procedure was used to extract unit cell parameters \cite{pawley_unit-cell_1981}. The refined unit cell volumes are in good agreement with prior structural studies on bulk LaH$_{10}$ samples in Refs. \cite{drozdov_superconductivity_2019, geballe_synthesis_2018, somayazulu_evidence_2019, laniel_high-pressure_2022, sun_high-temperature_2021} as demonstrated in Figure \ref{fig:XRD_summary}(c). Since only the $fcc$ lanthanum sublattice is detected in diffraction, there is an uncertainty in inferring the hydrogen stoichiometry from the measured unit cell volume, and we cannot rule out that there is a variation in hydrogen content between our samples and those synthesised in different studies via different formation pathways. We therefore refer to the synthesised hydride as LaH$_{10\pm\delta}$, where $\delta$ represents the deviation from the ideal LaH$_{10}$ stoichiometry. 


In order to assign the electrical properties of the hydride films to the crystal structures present within 
the sample, high-resolution spatial X-ray diffraction mapping was performed for both DACs on the ID27 beamline at the 
ESRF as shown in Figures \ref{fig:XRD_summary}(d) and (e). Spatial maps for both DACs were performed over a \SI{25}
{}$\times$\SI{25}{\micro\metre\squared} grid, with \SI{1}{\micro\metre} step size and with a beam diameter of \SI{0.7}
{\micro\metre} at FWHM. Photomicrographs of each sample after laser heating are shown below each respective spatial X-ray mapping. In both DACs, the intense (111) reflection corresponding to elemental Au was observed, which traced 
clearly the five electrodes as shown in Figures \ref{fig:XRD_summary}(d) and (e). After one laser-heating cycle, the 
film in DAC 1 consisted of a phase mixture of cubic $Fm\bar3m$-LaH$_{10\pm\delta}$ in the laser-heated region, as 
shown by the blue region in Figure \ref{fig:XRD_summary}(d), with highly localised regions of hexagonal $P6_3/mmc$-
LaH$_{9\pm\delta}$ on the boundary of the $Fm\bar3m$ phase (black regions). Hexagonal LaH$_{9\pm\delta}$ impurities 
have been observed in almost all prior LaH$_{10}$ syntheses reported in the literature 
\cite{drozdov_superconductivity_2019, sun_high-temperature_2021, laniel_high-pressure_2022}, and is consistent with theoretical structure searches that predict the stability of this phase under similar conditions \cite{shipley_stability_2020, kruglov_superconductivity_2020}. Further away from the laser-heated region we identified body-centred tetragonal $I4/mmm$-LaH$_{4\pm\delta}$, shown as the red region in 
Figure \ref{fig:XRD_summary}(d). We note that the spatial maps were constructed using the intensity of the LaH$_{10\pm\delta}$ (311) reflection and the LaH$_{9\pm\delta}$ (101) and 
LaH$_{4\pm\delta}$ (101) reflections due to their separated 2$\theta$ positions, enabling unambiguous mapping of the phase 
distributions. The coexistence of $I4/mmm$, $P6_3/mmc$, and $Fm\bar3m$ hydrides at the synthesis pressure of DAC 1 is 
in agreement with the findings of Laniel \textit{et al.,} in Ref. \cite{laniel_high-pressure_2022}. In both DACs, a diffraction peak was observed at $2\theta=\SI{11.2}{\degree}$ which appeared to be localised to the dark laser-heated region on each sample, visible in the integrated pattern for DAC 1 in Fig. \ref{fig:XRD_summary}(a). We cannot explain this peak with elemental lanthanum or any of the known lanthanum hydride phases \cite{laniel_high-pressure_2022}, but we do find that the peak is consistent with the (111) reflection of $F\bar43m$ cubic boron nitride (\textit{c}-BN) at this pressure \cite{datchi_equation_2007} which could explain the origin of the dark region on each sample. It is plausible that \textit{c}-BN formed directly above the sample on laser heating via the complete decomposition of ammonia borane where the highest temperatures were achieved, although we acknowledge that this is a tentative assignment. Two further reflections observed in DAC 1 at $2\theta= \SI{19.6}{\degree}$ and \SI{22.9}{\degree} could not be indexed, and are indicated with an asterisk in Figure \ref{fig:XRD_summary}(a).

The decrease in hydrogen stoichiometry with increasing distance from the laser-heated region likely arises from a lateral thermal gradient across the La film during laser heating, and hence a variation of hydrogen excess across the sample. Ammonia borane (AB) undergoes a multistep decomposition and hydrogen release as temperature increases \cite{demirci_ammonia_2017}. Consequently, hydrogen release from the surrounding AB during laser heating is likely to be spatially non-uniform due to these temperature gradients across the sample. 
Interestingly, Raman spectroscopy mapping of the 
molecular H$_2$ vibron signal across the sample in DAC 1 (measured 146 days after laser heating, 
Fig.~\ref{fig: DAC 1_Pressure_Vibron}) reveals that free molecular hydrogen, 
identified by its characteristic vibron, remained present in the region above the 
LaH$_{10\pm\delta}$ phase likely where the highest temperature and greatest hydrogen excess were achieved during laser 
heating. The H$_2$ vibron remained present after 301 and 370 days post-synthesis in DAC 
1 and 2 respectively (Figs. \ref{fig:DAC 1_Raman stability} and \ref{fig: DAC 2_Pressure Raman}), suggesting that hydrogen is trapped in a region above the LaH$_{10\pm\delta}$ phase and surrounded by partially decomposed ammonia borane, and does not diffuse out of the sample chamber, e.g. through the diamond anvils or the gasket, over such timescales. Further laser heating at various positions over the film would likely have transformed LaH$_{4\pm\delta}$ to LaH$_{10\pm\delta}$, although this was not done for DAC 1 to reduce the risk of anvil failure before electrical measurements could be carried out. 

Two laser-heating cycles were performed for the film in DAC 2, which resulted in an almost complete transformation to the $Fm\bar3m$-LaH$_{10\pm\delta}$ phase, which made contact with all five electrodes as shown in Figure \ref{fig:XRD_summary}(e). Localised regions of $P6_3/mmc$-LaH$_{9\pm\delta}$ on the outer boundary of the $Fm\bar3m$ phase were also observed, as in DAC 1. No evidence of $I4/mmm$-LaH$_{4\pm\delta}$ was observed in DAC 2 confirming sufficient hydrogen excess around the sample during laser heating. Representative integrated patterns demonstrating the presence of each phase within the DACs are shown in supplementary Figures \ref{fig: DAC 1_representative frames} and \ref{fig: DAC 2_hcp LaH9}.

\subsection{Electrical transport measurements}

\begin{figure}[ht!]
    \centering
    \includegraphics[width=0.9\textwidth]{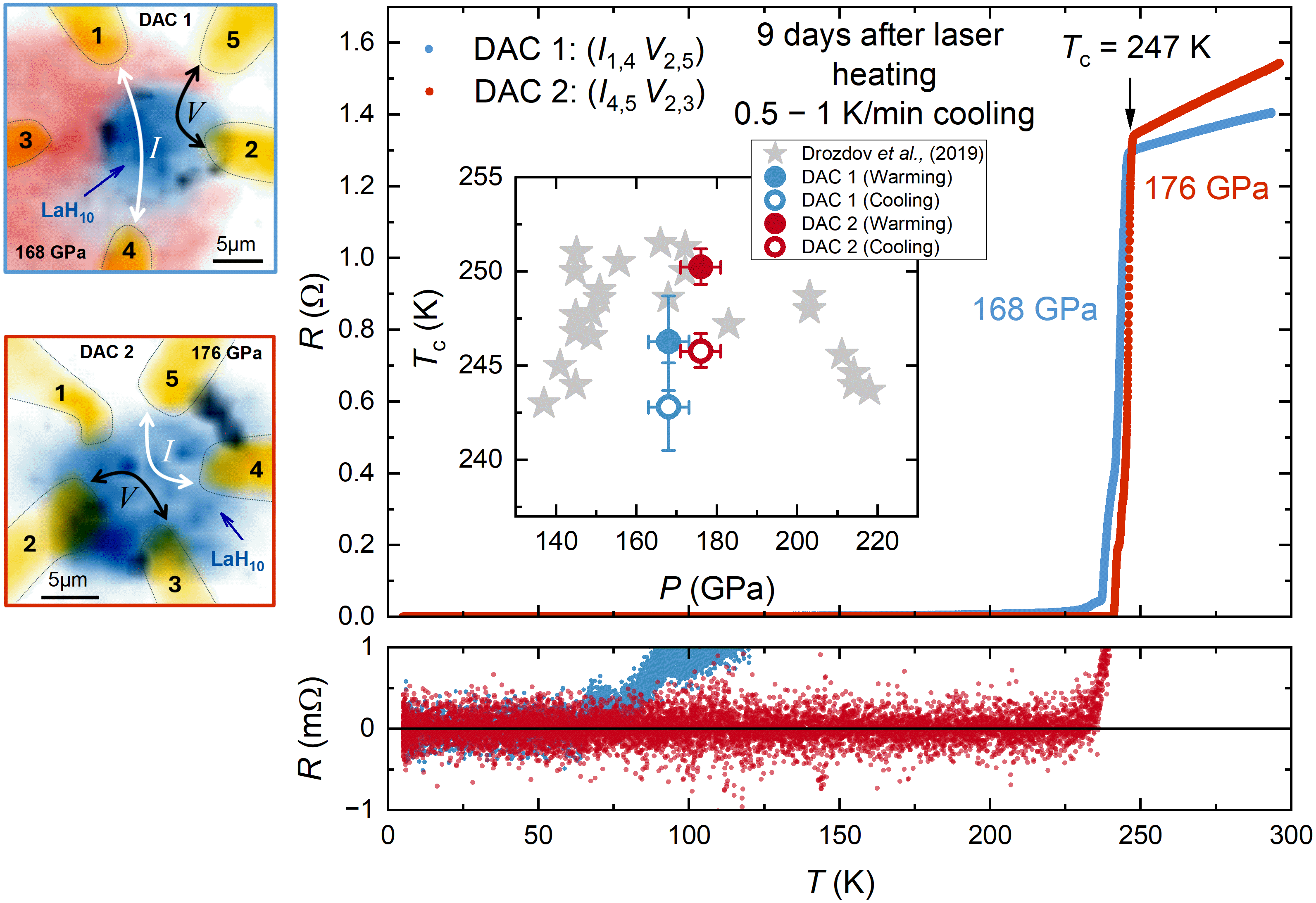}
    \caption{Four-point resistance versus temperature measured for both DACs 9 days after their respective laser heating. Measurements were taken in cooling cycles with cooling rate \SI{0.5}{}-\SI{1}{\kelvin\per\minute}, and with excitation current of \SI{30}{\micro\ampere}. Lower panel shows the zero resistance region. Inset shows $T_c$ versus pressure to compare $T_c$ of the films with the reported values for bulk LaH$_{10}$ in Ref. \cite{drozdov_superconductivity_2019}. Spatial XRD mappings for each DAC to the left of the main panel indicate the electrode configuration used to measure each curve.
    }
    \label{fig:RTs_bothDACs_main}
\end{figure}

Prior to laser heating, the metallic character of the elemental lanthanum films was verified in measurements of resistance versus temperature (Fig. \ref{fig:RTs La before LH_SI}). After laser heating, the normal state resistances increased for both DACs, and the onset of resistive transitions were observed at $T_c=\SI{245}{\kelvin}$ and $T_c=\SI{247}{\kelvin}$ in DACs 1 and 2 respectively as shown in Figure \ref{fig:RTs_bothDACs_main}, as extracted from cooling cycles at rates of between \SI{0.5}{}-\SI{1}{\kelvin\per\minute}. A careful analysis of the extrinsic thermal hysteresis observed in our measurement systems is provided in Fig. \ref{fig:RT_hysteresis}. The observed values of $T_c$ are in good agreement with those reported in the literature for bulk LaH$_{10}$, shown in the inset of Figure \ref{fig:RTs_bothDACs_main}, and with the value predicted for the $Fm\bar3m$-LaH$_{10}$ structure in theoretical studies \cite{errea_quantum_2020}. 
In DAC 1, there is a sharp resistance drop by a factor of 28 at \SI{245}{\kelvin}, with a residual normal state contribution attributed to $I4/mmm$-LaH$_{4\pm\delta}$ that was observed in the XRD measurements. There appears a second resistive transition in DAC 1 at $\sim \SI{70}{\kelvin}$ before the resistance falls to zero within the noise. A complete percolation pathway between the voltage probes necessary for zero resistance in DAC 1 is realised for the $I4/mmm$ phase but not for the localised regions of the $P6_3/mmc$ phase, evidenced in the XRD spatial maps in Fig. \ref{fig:XRD_summary}(d). We therefore tentatively assign this lower transition to the $I4/mmm$ phase, further supported by a recent study reporting a higher $T_c$ for the $P6_3/mmc$ phase, at around \SI{160}{\kelvin} at \SI{140}{\GPa} \cite{gorelli_raman_2025}. Further targeted syntheses of $I4/mmm$-LaH$_{4\pm\delta}$ would be desirable. The electrode configuration used to measure the curve for DAC 1 in Figure \ref{fig:RTs_bothDACs_main} was chosen such that the voltage probes were closest to the $Fm\bar3m$-LaH$_{10\pm\delta}$ phase. Measurements in various contact configurations for DAC 1 are shown in Figure \ref{fig: Config summary}(a). 

Following the first laser heating of DAC 2, a broad transition to a zero-resistance state was observed with an onset at $T_c=\SI{244}{\kelvin}$ and transition width ($ \Delta T_c=T_{c}^{onset}-T_{c}^{zero}$) of about \SI{30}{\kelvin} (red curve in Fig. \ref{fig: Config summary}(b)). After the second laser heating of DAC 2, the normal state resistance increased further and a sharp resistance drop by a factor $> 10^4$ was observed at $T_c=\SI{247}{\kelvin}$ as shown by the red curve in Figure \ref{fig:RTs_bothDACs_main}. The transition width was about \SI{12}{\kelvin} ($ \Delta T_c/T_c \sim 5\%$). We attribute the further increase in the normal state resistance to the transformation of a larger volume fraction of the elemental lanthanum film to the $Fm\bar3m$-LaH$_{10\pm\delta}$ phase on successive laser heating cycles.
After the second laser heating of DAC 2, the resistance reaches zero, within the noise of the measurement, below \SI{235}{\kelvin}. Zero resistance is observed in all measured electrode configurations of DAC 2 (Figs. \ref{fig: Config summary}(b)--(d)), consistent with the $Fm\bar3m$ phase forming a continuous pathway between all electrodes as evidenced by the XRD spatial map in Figure \ref{fig:XRD_summary}(e).


In the zero-resistance regime, the measured RMS resistance noise with a \SI{30}{\micro\ampere} excitation current was found to be 
$\sigma=\SI{64.3}{\micro\ohm}$ and $\sigma=\SI{74.3}{\micro\ohm}$ for DAC~1 and DAC~2, respectively (Figs.~\ref{fig: Noise analysis_all DACs}(a) and (b)). This is 
consistent with the dominant input noise of the SRS SIM921 preamplifier (JFET LSK489), which is $\SI{60}{\micro\ohm}$ for a \SI{30}{\micro\ampere} excitation and with the \SI{1}{\second} time constant used in our measurements. We compare this noise floor for DAC 2 with the expected resistance of a high-purity copper sample with residual resistivity ratio RRR $\approx 5760$ \cite{li_thermal_1988}, with the same dimensions, providing a stringent benchmark for the zero-resistance state. The copper resistance is estimated from its known resistivity, $\rho$ \cite{li_thermal_1988}, 
and the measured dimensions of the sample in DAC 2 (length $l=\SI{3.9}{\micro\metre}$, width $w=\SI{26}{\micro\metre}$, thickness $t=\SI{213.5}{\nano\metre}$) according to $R=\rho l/wt$. To obtain a conservative lower-bound estimate of the copper resistance, we use the deposited lanthanum thickness at ambient pressure and neglect the reduction in sample thickness under pressure. We find that at \SI{235}{\kelvin}, just below $T_c$ where the measured resistance reaches zero within the noise floor, the expected resistance of copper is approximately \SI{9}{\milli\ohm}. Our noise floor is therefore more than two orders of magnitude smaller than the resistance of copper at this temperature (Fig.~\ref{fig: Noise analysis_all DACs}(c)), and demonstrates that our measurement resolution is sufficient to clearly distinguish the superconducting zero-resistance state from the lowest resistances achieved in highest purity copper. Below approximately \SI{35}{\kelvin} however, the residual resistance of copper ($\sim\SI{2}{\micro\ohm}$) falls below our noise floor (inset of Fig.~\ref{fig: Noise analysis_all DACs}(c)). This illustrates that demonstrating a zero-resistance state at lowest temperature with a resistance resolution below the 
residual resistance of even the highest-purity metals remains experimentally challenging.
In principle, the resistance resolution could be improved by applying an excitation current 50--100 times larger. However, while increasing the 
excitation current improves the voltage sensitivity, it could also introduce systematic effects associated with the high current density, including common-mode leakage due to the finite contact resistance of the current leads.

\subsection{Upper critical field $H_{c2}(T)$}

\begin{figure}[ht!]
    \centering
    \includegraphics[width=\textwidth]{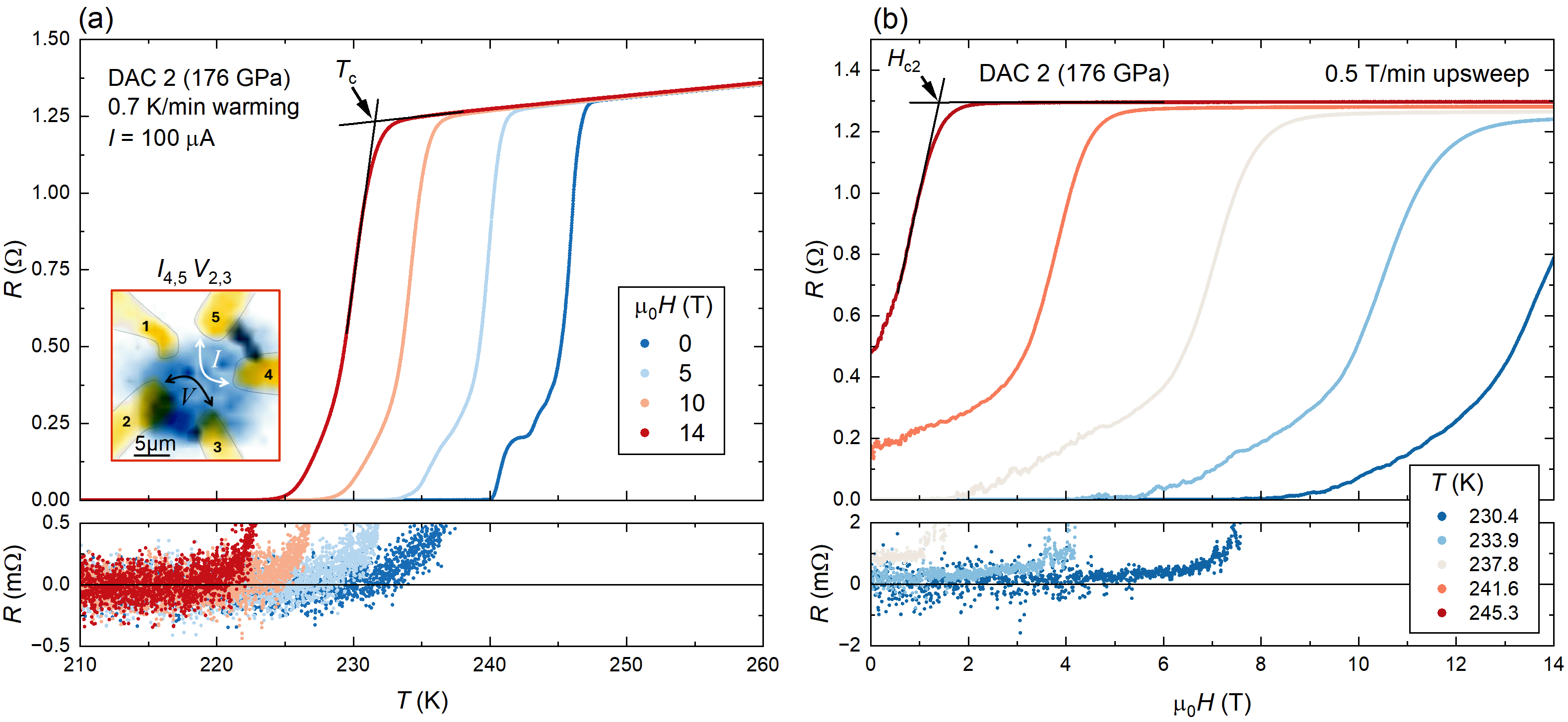}
    \caption{Suppression of the superconducting transition of the LaH$_{10 \pm\delta }$ film in DAC 2 at \SI{176}{\giga\pascal}. (a) Temperature sweeps in static magnetic fields up to \SI{14}{\tesla}, demonstrating the downward shift in critical temperature $T_c$. The value of $T_c$ was extracted as the intersection between linear fits to the normal state and superconducting transition as indicated for the \SI{14}{\tesla} curve. The XRD mapping of the $Fm\bar3m$ phase is shown in the inset with the electrode configuration for the measurements. (b) Isothermal field sweeps for the sample in DAC 2. The value of $H_{c2}$ was extracted as indicated for the curve measured at $T = \SI{245.3}{\kelvin}$. Measurements of DAC 1 in magnetic field are shown in Figure \ref{fig: Hc2_DAC1_SI}.
    }
    \label{fig:Hc2_DAC2}
\end{figure}

\begin{figure}[ht!]
    \centering
    \includegraphics[width=0.7\textwidth]{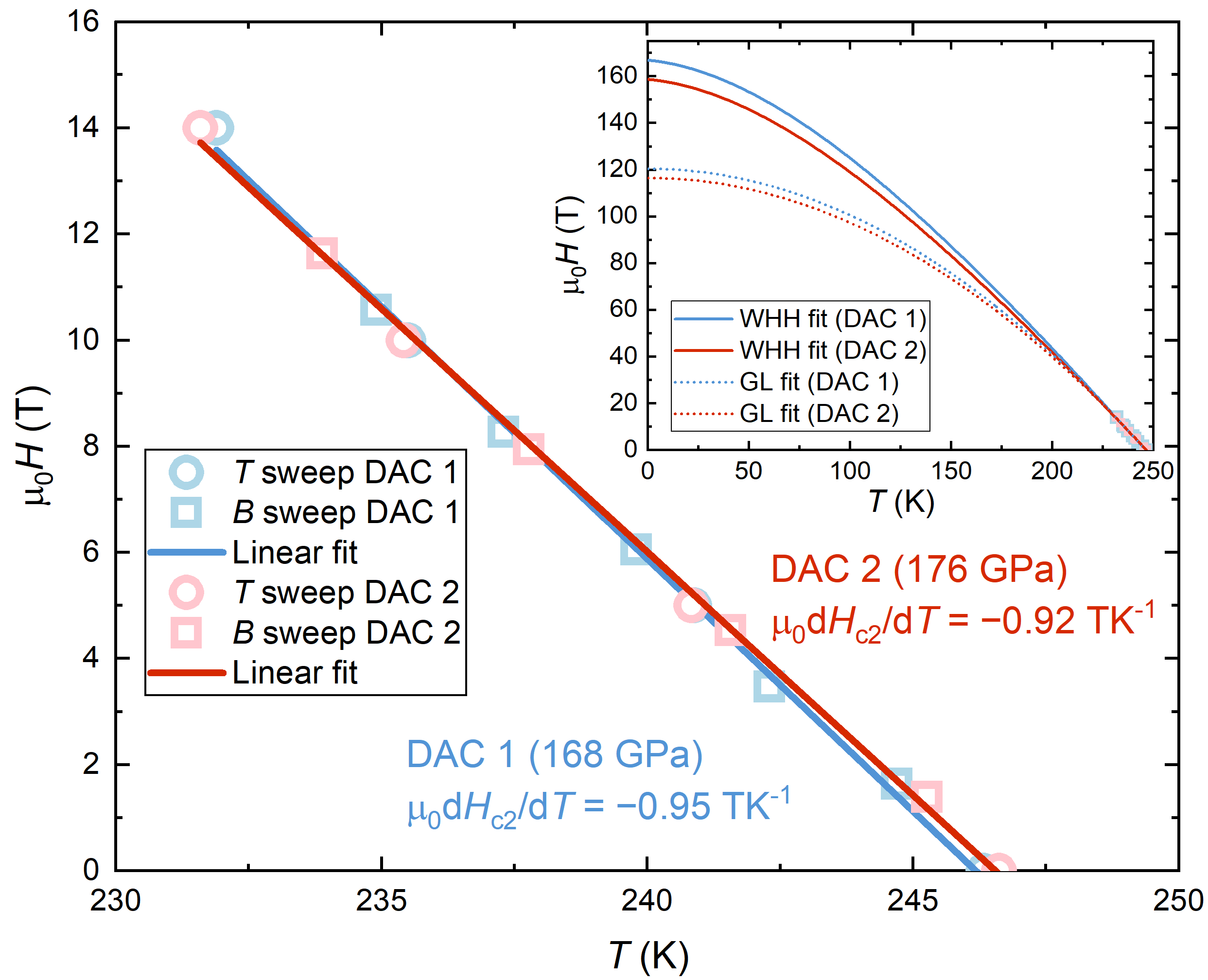}
    \caption{The magnetic phase diagram near $T_c$ for the LaH$_{10 \pm\delta}$ films constructed from measurements of $T_c(\mu_0H)$ and $\mu_0H_{c2}(T)$ in Figures \ref{fig:Hc2_DAC2} and \ref{fig: Hc2_DAC1_SI}, demonstrating linear behaviour of $\mu_0H_{c2}(T)$ close to $T_c$. The slope near $T_c$, $\mu_0$d$H_{c2}$/d$T$$|_{T_c}$, is indicated for each DAC. Inset shows extrapolated fits to the Ginzburg-Landau (GL) and simplified Werthamer–Helfand–Hohenberg (WHH) models to estimate $\mu_0H_{c2}(0)$, found to be in the range \SI{120}{}-\SI{167}{\tesla} for DAC 1 and \SI{116}{}-\SI{159}{\tesla} for DAC 2.
    }
    \label{fig:Hc2_figure_nearTc}
\end{figure}

In order to provide further evidence for superconductivity in the LaH$_{10 \pm\delta }$ films, resistance measurements were performed in magnetic fields up to \SI{14}{\tesla}. Temperature sweeps in static field and isothermal field sweeps as measured in DAC 2 are shown in Figures \ref{fig:Hc2_DAC2}(a) and (b) respectively. The transition temperature was suppressed by about \SI{15}{\kelvin} at \SI{14}{\tesla} as shown in Figure \ref{fig:Hc2_DAC2}(a). We observe a broad step at around \SI{240}{\kelvin} in the zero-field measurement in Figure \ref{fig:Hc2_DAC2}(a), which disappears above \SI{5}{\tesla}. Very similar behaviour was observed in a bulk sample by Drozdov \textit{et al.,} \cite{drozdov_superconductivity_2019}, and we attribute this to inhomogeneity of the superconducting phase at this temperature. Measurements performed in magnetic field for DAC 1 are shown in Figure \ref{fig: Hc2_DAC1_SI}. The magnetic phase diagram is shown in Figure \ref{fig:Hc2_figure_nearTc}, constructed by extracting $T_c(H)$ and $H_{c2}(T)$ from the intersection of linear fits to the normal state and superconducting transitions for both samples as shown in Figure \ref{fig:Hc2_DAC2}. Linear behaviour of $H_{c2}(T)$ is observed close to $T_c$, with the slope d$H_{c2}/$d$T$ equal to \SI{-0.95}{\tesla\per\kelvin} and \SI{-0.92}{\tesla\per\kelvin} in DACs 1 and 2 respectively, close to values reported in Refs. \cite{drozdov_superconductivity_2019, sun_high-temperature_2021}. Fits to the Ginzburg-Landau (GL) and simplified Werthamer–Helfand–Hohenberg (WHH) models \cite{baumgartner_effects_2013} yield estimates of the zero-temperature limit of the upper critical field $H_{c2}(0)$ in the range $\SI{120}{}-\SI{167}{\tesla}$ for DAC 1 and $\SI{116}{}-\SI{159}{\tesla}$ for DAC 2 respectively, in good agreement with previous studies of bulk LaH$_{10}$ samples \cite{drozdov_superconductivity_2019, sun_high-temperature_2021}. The best estimate of $H_{c2}(0)$ for $Fm\bar3m$-LaH$_{10}$ is around \SI{144}{\tesla} as determined by Sun \textit{et al.,} from WHH fits to pulsed field measurements up to \SI{60}{\tesla} \cite{sun_high-temperature_2021}. We note that extrapolations of GL and WHH models may not capture any strong coupling or multiband effects that could become prominent towards lower temperatures in $H_{c2}(T)$, and that direct measurement of $H_{c2}(0)$ requiring magnetic fields in excess of \SI{100}{\tesla} is beyond current experimental capabilities for samples within a DAC. 

Estimates of $H_{c2}(0)$ from the WHH model are about a factor of 3 smaller than that expected from the weak-coupling Pauli limiting field ($H_p(0)\sim1.85 T_c \sim \SI{455}{\tesla}$ \cite{clogston_upper_1962}) suggesting orbital pair-breaking effects are dominant in our LaH$_{10\pm\delta}$ films, consistent with the findings for bulk LaH$_{10}$ \cite{drozdov_superconductivity_2019, sun_high-temperature_2021} as well as other high-$T_c$ hydrides measured to date \cite{cross_high-temperature_2024,osmond_clean-limit_2022, mozaffari_superconducting_2019, kong_superconductivity_2021, chen_high-temperature_2021}. In line with previous measurements on bulk samples, we estimate the superconducting coherence length $\xi(0)$ according to the Ginzburg-Landau equation $\mu_0H_{c2}(0)=\Phi_0/2\pi\xi(0)^2$ for an orbital-limited type-II superconductor, where $\Phi_0=h/2e$ is the magnetic flux quantum. Substituting $H_{c2}(0)$ as estimated from GL and WHH models, we find $\xi(0)=\SI{1.40}{}-\SI{1.66}{\nano\metre}$ for DAC 1, and $\xi(0)=\SI{1.44}{}-\SI{1.68}{\nano\metre}$ for DAC 2, in line with values reported in previous studies \cite{drozdov_superconductivity_2019, sun_high-temperature_2021}. Differences observed in the coherence length and the slope near $T_c$ likely arise due to differences in sample purity and mean free path between 
the different studies. 
A rigorous classification of LaH$_{10}$ as a type-II superconductor would require knowledge of both the superconducting coherence length, $\xi$, and the magnetic penetration depth, $\lambda$, through the Ginzburg-Landau parameter $\kappa=\lambda/\xi$. Since $\lambda$ is not measured in the present work, the coherence length alone is insufficient to establish $\kappa > 1/\sqrt{2}$, which defines a type-II superconductor. Nevertheless, the short coherence lengths obtained here are commonly associated with strongly type-II superconductivity, and indeed they are comparable to those of type-II unconventional superconductors such as Bi-2223 ($\xi_{ab} \sim$\SI{1.5}{\nano\metre}) and smaller than those of type-II conventional superconductors including MgB$_2$ ($\xi_{ab}\sim \SI{6.5}{\nano\metre}$) and Nb$_3$Sn ($\xi_{ab}\sim \SI{3}{\nano\metre}$) \cite{yao_superconducting_2021}. From the inferred values of $\xi(0)$, we estimate the weak-coupling BCS Fermi velocity $v_F$ using $\xi(0)=\hbar v_F/1.76\pi k_B T_c$. Using the upper and lower values of $\xi(0)$, $v_F$ is found to be in the range $\SI{2.50}{}-\SI{2.96e5}{\metre\per\second}$ and $\SI{2.57}{}-\SI{3.00e5}{\metre\per\second}$ in DACs 1 and 2 respectively, within the range of universal Fermi velocities reported in Refs. \cite{sun_high-temperature_2021, talantsev_universal_2022} and in good agreement with previous reports. A summary of the superconducting properties of LaH$_{10\pm\delta}$ found in this work and in prior studies is given in Table \ref{table:superconducting_parameters}.

\begin{table}[ht]
\caption{Superconducting parameters of $Fm\bar3m$-LaH$_{10\pm\delta}$ from different high-pressure studies. Samples were synthesised from La films + AB in this work, while works of Drozdov \textit{et al.,} and Sun \textit{et al.,} used bulk La and pure H$_2$ \cite{drozdov_superconductivity_2019, sun_high-temperature_2021}. The values of $T_c$ displayed in the table for this work are those extracted from warming curves at \SI{0.7}{\kelvin \per \minute}. The lower and upper values given for $\mu_0H_{c2}(0)$ in this work are those extracted from GL and WHH extrapolations respectively, while those from Drozdov \textit{et al.,} were obtained from GL fits to $H_{c2}(T)$ constructed using $T_c$ values defined by the 90\% and 50\% normal-state resistance criteria reported in Ref. \cite{drozdov_superconductivity_2019}. The range of values for $\xi(0)$ and BCS $v_F$ in each study are calculated from the corresponding range of $\mu_0H_{c2}(0)$ values according to the equations given in the main text.} 

\centering
\setlength{\tabcolsep}{5pt} 
\renewcommand{\arraystretch}{1.2}

\begin{tabular}{|l|c|c|c|c|c|c|}
\hline
\multicolumn{1}{|c|}{Study} &
$P$ &
$T_c$ &
$\mu_0\left.\dfrac{\mathrm{d}H_{c2}}{\mathrm{d}T}\right|_{T_c}$ &
$\mu_0 H_{c2}(0)$ &
$\xi(0)$ &
BCS $v_F$
\\
&
(GPa) &
(K) &
(\SI{}{\tesla\per\kelvin}) &
(T) &
(nm) &
($\times 10^5$ \SI{}{\metre\per\second}) \\
\hline

DAC 1 (this work)
& $168 \pm 5$
& 246.3
& --0.95
& 120--167
& 1.40--1.66
& 2.50--2.96 
\\

DAC 2 (this work)
& $176 \pm 5$
& 246.6
& --0.92
& 116--159
& 1.44--1.68
& 2.57--3.00
\\

Drozdov \textit{et al.,}~\cite{drozdov_superconductivity_2019}
& 150
& 249
& --0.90
& 95--136
& 1.56--1.86
& 2.81--3.35
\\

Sun \textit{et al.,}~\cite{sun_high-temperature_2021}
& 136
& 246
& --0.83
& 143.5
& 1.51
& 2.77
\\

\hline
\end{tabular}

\label{table:superconducting_parameters}

\end{table}

For type-II superconductors, a broadening of the resistive transition is expected in magnetic field due to dissipation associated with the motion of vortices in the mixed state \cite{talantsev2024comment, hirsch_nonstandard_2021}.
The apparent field dependence of the transition width can, however, depend sensitively on the criterion used to define the superconducting transition, particularly in the presence of additional resistive features. Using the conventional 90\% and 10\% resistance criteria \cite{talantsev2024comment}, the transition widths of our LaH$_{10\pm\delta}$ films initially decrease with magnetic field before gradually increasing above \SI{14}{\tesla} (see Fig. \ref{fig: Tc Broadening in field}(a)). Qualitatively similar behaviour displaying an initial narrowing of the resistive transition width followed by broadening toward higher fields was shown across various classes of both conventional and unconventional type-II superconductors in Ref. \cite{talantsev2024comment}, including in H$_3$S, MgB$_2$ thin films, pnictides and cuprates. In this work, the initial narrowing arises primarily from the field suppression of the broad resistive step near 240 K, which we attribute to inhomogeneity of the superconducting phase (see Fig. \ref{fig: Tc Broadening in field}(b)). When the transition width is instead defined using the 90\% and 50\% resistance criteria, similar to that employed in Ref. \cite{osmond_clean-limit_2022}, thereby excluding this feature, it increases monotonically with magnetic field (see Fig. \ref{fig: Tc Broadening in field} (a)). Since the maximum applied field of 14 T corresponds to less than 10\% of the extrapolated $\mu_0H_{c2}(0)$, more pronounced broadening would be expected at higher magnetic fields.


\section{Temporal stability of the crystal structure and superconductivity of LaH$_{10\pm\delta}$ films}
\label{sec:Temporal stability}

One important question to address is the long-term stability of lanthanum hydrides over time, and whether hydrogen diffusion drives any temporal electronic or structural changes. Such phenomena have been observed previously in hydride compounds, for example in the high-$T_c$ hydride $Pm\bar3n$-La$_4$H$_{23}$ whereby it was proposed that hydrogen diffusion could be driving observed increases in the normal state resistance at temperatures above \SI{250}{\kelvin} \cite{cross_high-temperature_2024}. Further insight into the effects of hydrogen vacancies in La$_4$H$_{23}$ has recently been presented in Ref. \cite{osmond_hydrogen_2026}. There are also open questions about the long-term stability of $Fm\bar3m$-LaH$_{10}$. Recent NMR measurements have shown evidence of a continuous dehydrogenation of LaH$_{10+\delta}$ ($\delta = 0.2-1.1$) to a stoichiometry close to that of the precursor trihydride LaH$_3$ over the course of about 70 days after laser heating and at constant pressure, with a consistent dehydrogenation timescale measured across four DACs \cite{zhou_diffusion-driven_2025}. The observed dehydrogenation was claimed to be further supported by electrical measurements that showed a tendency for $T_c$ to decrease over time in four DACs containing lanthanum hydride compounds.

\begin{figure}[ht!]
    \centering
    \includegraphics[width=\textwidth]{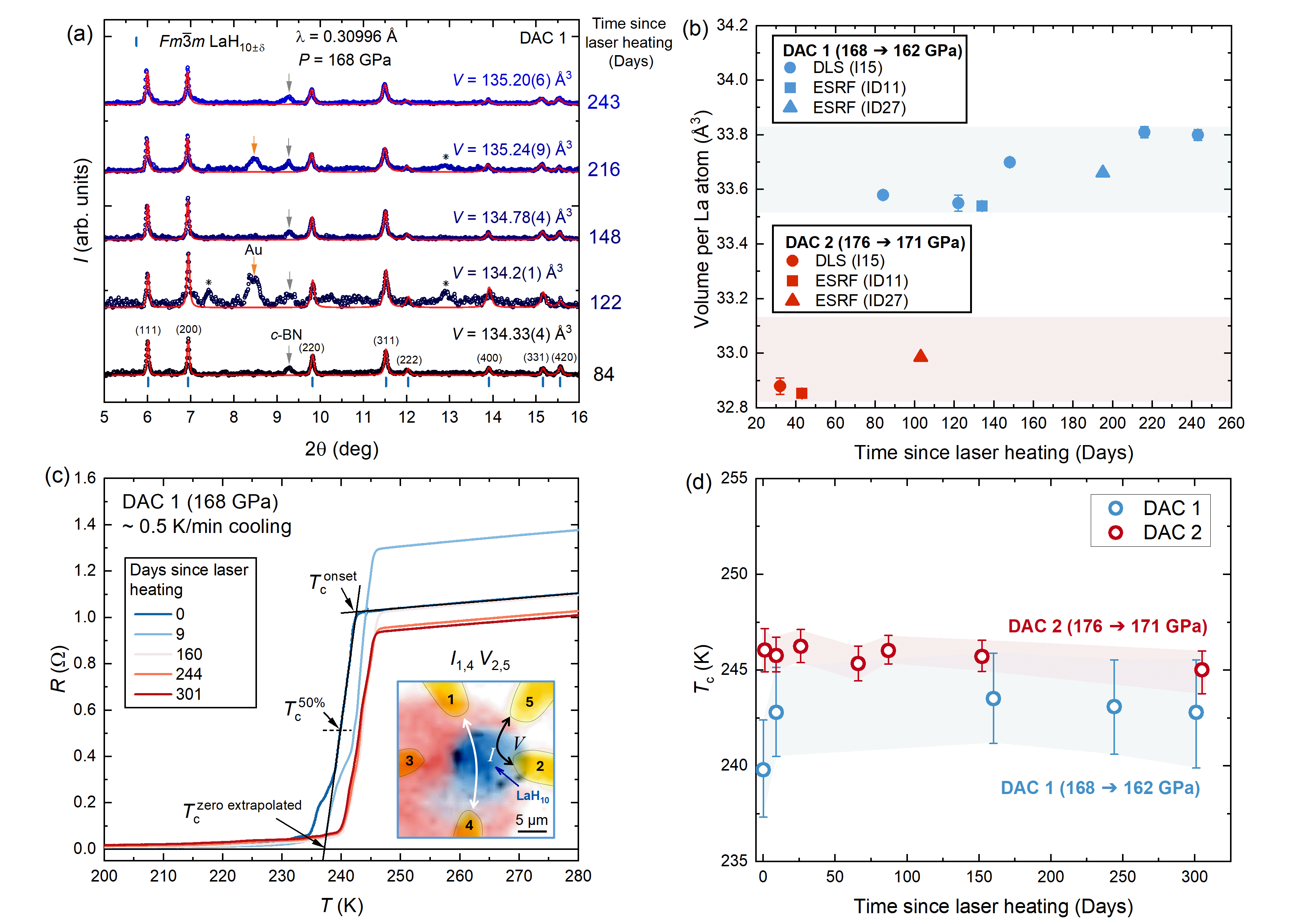}
    \caption{Stability of LaH$_{10 \pm \delta}$ films over time. (a) XRD data and structural refinements to the $Fm\bar3m$ phase for DAC 1 at \SI{168}{\GPa}. Data are shown as open circles, with Pawley refinements shown as the red solid curves. All data were normalised to the intensity of the (111) reflection for comparison. Measurements were taken on the I15 beamline at Diamond light source over a period from 84 to 243 days after laser heating. The volumes obtained from refinements are displayed alongside their corresponding curves. Where present, peaks due to elemental Au from the electrodes and \textit{c}-BN from AB decomposition are indicated, and unassigned peaks are denoted with an asterisk ($\ast$). (b) Unit cell volume as a function of time after laser heating for both DACs, including data from beamlines I15 (see Fig. \ref{fig: DAC2_June DLS} for DAC 2 refinement), ID11 (Fig. \ref{fig: ID11_Summary}) and ID27 (Fig. \ref{fig:XRD_summary}(a)). Shaded bands indicate the change in volume expected from the DFT equation of state in Ref. \cite{laniel_high-pressure_2022}, given a \SI{5}{\GPa} pressure decrease in both DACs over the measurement period. (c) Resistance versus temperature curves for DAC 1 measured up to 301 days after laser-heating. (d) The stability of the superconducting transition temperature $T_c$ over time, extracted from cooling cycles. Data points correspond to $T_c$ extracted at 50\% ($T_c^{50\%}$) of the resistance drop. Asymmetric error bars reflect the difference between $T_c^{50\%}$ and $T_c^{onset}$ (upper error bar), and $T_c^{zero\,extrapolated}$ (lower error bar), as defined in (c).
    }
    \label{fig:Stability_multipanel}
\end{figure}

In contrast to the NMR study, Minkov \textit{et al.,} demonstrated the long-term stability of both the crystal structure and superconductivity of $Fm\bar3m$-LaH$_{10}$ using combined XRD and transport measurements \cite{minkov_long-term_2026}. The unit-cell volume was shown to remain stable for approximately 4.5 years, with a small decrease in volume of the order \SI{0.3}{\angstrom\cubed} per La atom attributed to a small pressure increase by about \SI{5}{\GPa}. Superconductivity with $T_c = \SI{247}{\kelvin}$ was shown to be stable over timescales exceeding 5 years. In order to further address the questions surrounding the stability of $Fm\bar3m$-LaH$_{10\pm\delta}$, we performed a combined high-resolution synchrotron XRD and electrical transport study of our high-$T_c$ films over a period of about 300 days post laser-heating synthesis, beyond the timescale over which LaH$_{10\pm\delta}$ was observed to decompose in NMR measurements \cite{zhou_diffusion-driven_2025}. A summary of our stability study is shown in Figure \ref{fig:Stability_multipanel}. 

Figure \ref{fig:Stability_multipanel}(a) shows XRD measurements performed on beamline I15 at Diamond Light Source for DAC 1 between 84 and 243 days post laser heating synthesis. Integrated patterns show clearly the presence of the $Fm\bar3m$ phase up to 243 days after laser heating. The unit cell volumes extracted from structural refinements are displayed alongside each integrated pattern in (a). Unit cell volume is shown as a function of time in Figure \ref{fig:Stability_multipanel}(b) demonstrating stability of the $Fm\bar3m$ crystal structure. Measurements presented in Figure \ref{fig:Stability_multipanel}(b) were conducted on three different beamlines: I15 (circles) (Fig. \ref{fig:Stability_multipanel}(a) and \ref{fig: DAC2_June DLS}), as well as on ESRF beamlines ID11 (squares) (see Fig. \ref{fig: ID11_Summary}) and ID27 (triangles) (see Fig. \ref{fig:XRD_summary}(a)). Further, we observe no significant changes in the spatial extents of the hydride phases in either DAC, evidenced by comparing the high-resolution XRD spatial maps performed on beamlines ID11 and ID27, acquired with a 60 day interval between the measurements (see Fig. \ref{fig: ESRF_Mapping comparison}). We emphasise that such a qualitative comparison does not rule out the possibility of the formation of an amorphous hydride phase over time, whether via reaction with surrounding free hydrogen or by decomposition of any of the identified hydride phases. Nevertheless, the retention of largely the same extent of $Fm\bar3m$-LaH$_{10\pm\delta}$ and the absence of signatures of $Fm\bar3m$-LaH$_3$ in any of our XRD measurements strongly imply that our high-$T_c$ films do not follow the decomposition pathway observed in the NMR study \cite{zhou_diffusion-driven_2025}. 


We observe small changes in the unit cell volume of the $Fm\bar3m$ phase over the course of the XRD measurements, with the largest variation of \SI{1.1}{\angstrom\cubed} in DAC 1 and \SI{0.5}{\angstrom\cubed} in DAC 2. This can very likely be attributed to small pressure decreases ($\sim 5 - \SI{6}{\GPa}$) in both DACs over the measurement period (Figs. \ref{fig:DAC 1_Raman stability} and \ref{fig: DAC 2_Pressure Raman}).  Using the DFT equation of state from Ref.~\cite{laniel_high-pressure_2022}, we estimate that the observed increase in unit cell volume can likely be accounted for with a pressure decrease of \SI{5}{\GPa} in each DAC, illustrated by the shaded bands in Fig. \ref{fig:Stability_multipanel}(b). We cannot rule out a further contribution to the volume increase due to hydrogen diffusion into the sample. We also note that integrated diffraction patterns acquired at the different beamtimes may not correspond exactly to the same position on the sample, and hence volumes may be slightly different due to pressure gradients in the DAC (see Figs. \ref{fig: DAC 1_Pressure_Vibron} and \ref{fig: DAC2_Pressure gradient}), or variations in hydrogen content across the sample. 

Figure \ref{fig:Stability_multipanel}(c) shows four-point resistance measurements of $T_c$ for DAC 1 from immediately after laser heating up to 301 days after synthesis, demonstrating stability of the superconducting transition. Measurements were taken during cooling at a rate of $\sim \SI{0.5}{\kelvin\per\minute}$ in the contact configuration as shown on the inset XRD spatial map. An overall decrease in the normal state resistance over time was observed for both DACs (see Fig. \ref{fig: DAC 2_RTs over time.png} for DAC 2 curves) while $T_c$ remained stable. Similar changes in the normal state were observed by Minkov \textit{et al.,} for bulk samples \cite{minkov_long-term_2026}, and we attribute this behaviour to annealing effects at room temperature. We believe this further rules out the decomposition to LaH$_3$, since this would be expected to drive an increase in the normal state resistance over time due to its semi-metallic nature at similar pressures \cite{drozdov_superconductivity_2019}. Figure \ref{fig:Stability_multipanel}(d) shows the measured $T_c$ values as a function of time for both DACs, extracted from cooling cycles. The value of $T_c$ in DAC 1 appears to show a slight increase in the first few days after laser-heating, but $T_c$ of both DACs is then stable over the course of about 300 days thereafter.

Our results reinforce the claims for the thermodynamic stability of LaH$_{10\pm\delta}$. This is indeed consistent with the findings in Refs. \cite{drozdov_superconductivity_2019, minkov_long-term_2025} whereby \textit{fcc} LaH$_{10}$ formed spontaneously without laser heating when LaH$_{3}$ was pressurised to \SI{140}{\GPa} in a molecular H$_2$ environment. This demonstrates that LaH$_{10}$ is more stable than LaH$_{3}$ at high pressures in excess H$_2$. One additional point to note is that in both this study and that of Minkov \textit{et al.,} the long-term retention of excess molecular hydrogen in the DACs after laser heating was reported (see Figs. \ref{fig:DAC 1_Raman stability}(c) and \ref{fig: DAC 2_Pressure Raman}(c) for this work). In the NMR study, no H$_2$ vibron was observed after laser heating \cite{zhou_diffusion-driven_2025}, and it was noted by the authors that the observed dehydrogenation of the hydride sample could be related to the absence of chemical equilibrium between the hydride and the surrounding hydrogen reservoir. This remains an important open question that warrants further investigation.
To further understand the decomposition observed in NMR measurements it would be desirable to perform combined XRD, Raman, and NMR measurements on samples synthesised via the same methods as in the NMR study, namely a laser-heated mixture of LaH$_3$ powder and ammonia borane. This would provide important insight into the synthesised crystal structures, reveal any sample heterogeneities and possible novel metastable phases, explore decomposition pathways and establish whether stabilisation requires a chemical equilibrium with excess molecular hydrogen in the sample chamber post laser heating. 

\section{Conclusion}
We demonstrate the first successful formation of $Fm\bar3m$-LaH$_{10\pm\delta}$ films in two diamond anvil cells (DACs) at pressures of \SI{168}{\GPa} and \SI{176}{\GPa}. The films were synthesised from an elemental lanthanum film deposited directly onto the anvils, and laser-heated at high pressures with purified ammonia borane as the hydrogen donor compound. High-resolution X-ray diffraction measurements confirmed the \textit{fcc} lanthanum sublattice of LaH$_{10\pm\delta}$, with unit cell parameters in excellent agreement with those reported for bulk samples in prior studies. Electrical transport measurements confirmed superconductivity with $T_c \sim \SI{247}{\kelvin}$, with the characteristic suppression observed in magnetic fields up to \SI{14}{\tesla}. Combined X-ray diffraction and electrical measurements reveal remarkable temporal stability of our films. We find no evidence of dehydrogenation during the 300-day period over which they were monitored following laser-heating, reinforcing the claims for the thermodynamic stability of $Fm\bar3m$-LaH$_{10\pm\delta}$ in an excess molecular H$_2$ environment. We further emphasise the importance of careful spatial diffraction mapping for reliable assignment of superconducting properties to the correct hydride phase, highly relevant for current ternary syntheses. Our work establishes thin-film methods as a promising platform for the future synthesis of ternary hydrides where accurate control of precursor stoichiometries will be critical. Thin-film techniques also provide a route toward the integration of micro-fabricated device geometries in diamond anvil cells for future detailed characterisation of the superconducting properties of hydrides.

\section{Experimental details}
\label{Experimental details}

The high-pressure synthesis of two LaH$_{10\pm\delta}$ films was carried out using custom DACs constructed from MP35N alloy, equipped with Type Ia Boehler \SI{30}{\degree} and Type IIas Boehler \SI{70}{\degree} (ultra-low fluorescence) anvils with culet diameters of \SI{50}{\micro\metre}, bevelled at \SI{8}{\degree} to a diameter of \SI{300}{\micro\metre}. Gaskets were prepared by pre-indentation of \SI{240}{\micro\metre} rhenium to a thickness of $\SI{50}{\micro\metre}$ followed by compression of an insulating boron nitride and epoxy mixture to a thickness of $\sim \SI{15}{\micro\metre}$. A hole of diameter \SI{40}{\micro\metre} was laser-drilled in the insulation forming the sample chamber. For electrical measurements, five tungsten-gold bi-layer electrodes were deposited through a shadow mask directly onto one of the diamond anvils, with the smallest separation between the tips of the electrodes $\sim \SI{4}{\micro\metre}$. Lanthanum films of thickness \SI{262}{\nano\metre} and \SI{213.5}{\nano\metre} were evaporated onto the electrodes in DACs 1 and 2 respectively at a rate of \SI{100}{\angstrom \per \second}, and in a vacuum of $\SI{4e-7}{\milli\bar}$ following multiple pre-melting cycles to remove oxide impurities from the source material. The DACs were loaded with ammonia borane (AB, NH$_3$BH$_3$) as the hydrogen donor and pressure transmitting medium (PTM). The AB was first purified by sublimation, and the DACs closed in an argon glovebox with residual O$_2$ and H$_2$O $<$ 0.5 ppm. DACs 1 and 2 were pressurised to \SI{172}{\GPa} and \SI{177}{\giga\pascal} respectively, with the pressure determined from the Raman shift of the high-frequency diamond edge using the Akahama calibration \cite{akahama_pressure_2006} (see Figs. \ref{fig: DAC 2_Pressure Raman} and \ref{fig: DAC1_Raman pressure + vibron}). Pressures determined from the position of the diamond edge typically yielded values \SI{5}{}--\SI{15}{\GPa} higher than those inferred from the hydrogen vibron. Prior to laser heating, the metallic character of the elemental lanthanum films was verified in resistance versus temperature measurements (Fig. \ref{fig:RTs La before LH_SI}), with residual resistance ratios (RRR) of 1.8 and 1.5 in DACs 1 and 2 respectively. The DACs were heated using a \SI{1070}{\nano\metre} Yb-fiber laser in a one-sided geometry \cite{lord_nisi_2014} with \SI{0.2}{}--\SI{0.3}{\second} pulses. The four-point resistance of each sample was monitored during laser heating. Laser power was incremented after each pulse until an irreversible increase in the room temperature resistance was observed, after which laser heating was halted. This resistance increase indicated a chemical reaction between the lanthanum films and the hydrogen released from the ammonia borane, and occurred at a temperature below that necessary to observe any glowing. Hence, we could not perform a reliable temperature measurement using black-body radiation during laser heating. Resistance versus temperature measurements were performed using a low-frequency AC technique in a four-point configuration with excitation currents in the range \SI{30}{\micro\ampere}--\SI{100}{\micro\ampere}. Zero-field measurements were done using an SRS SIM921 AC resistance bridge ($f \sim \SI{9}{\hertz}$), and magnetic field measurements were performed in a Cryogenic Ltd \SI{14}{\tesla} magnet using a Keithley 6221 current source, a model SR560 preamplifier and SR830 lock-in amplifier ($f \sim 16-\SI{18}{\hertz}$) with an excitation current of \SI{100}{\micro\ampere}. Resistances are presented as calculated from the in-phase component of the measured signal, with both the in-phase and quadrature components for each transport measurement provided in the data repository. X-ray diffraction measurements were performed on three beamlines: I15 beamline at Diamond Light Source, and ESRF beamlines ID11 and ID27. Details such as X-ray wavelength, beam diameter, detector model and sample-to-detector distance for measurements performed on each beamline are provided in the Supplementary Information Table \ref{table:xrd_setup}. Integrated diffraction patterns and azimuthal projections of the detector images were produced using DIOPTAS \cite{prescher_dioptas_2015} after masking of spurious reflections, and those from the diamond anvils. Background subtraction of the integrated patterns and subsequent structural refinements were performed using GSAS-II \cite{toby_gsas-ii_2013}. Spatial mapping of unique reflections corresponding to hydride phases within each DAC was performed using XDI \cite{hrubiak_multimode_2019}.

\section{Acknowledgements}

This work was supported by the EPSRC Open Fellowship of J.B. under Grant No. EP/Z533555/1. S.F. acknowledges support from the ERC Horizon 2020 programme under grant 715262-HPSuper and the EPSRC under grant EP/V048759/1. D.L. thanks the UKRI Future Leaders Fellowship (MR/V025724/1) for financial support. O.T.L. would like to acknowledge support from the Royal Society in the form of a University Research Fellowship (UF150057). We acknowledge Diamond Light Source for time on beamline I15 (proposals CY40128-2, CY40128-3, CY40128-4, CY40128-6, and CY40128-7). We acknowledge the European Synchrotron Radiation Facility (ESRF) for provision of synchrotron radiation facilities on beamlines ID11 and ID27 under proposals MA-6519 and HC-6190, respectively. We thank Jonathan Wright for assistance and support during the beamtime on ID11. We are grateful to Antony Carrington for useful discussions.

\section{Data availability}

Data will be made available at the University of Bristol data repository. X-ray diffraction data from ESRF beamlines ID11 (proposal: MA-6519) and ID27 (proposal: HC-6190) are available at DOI: 10.15151/ESRF-ES-2156316192 and DOI: 10.15151/ESRF-ES-2125107364, respectively.

\bibliography{bibliography}

\clearpage

\begin{center}
\Large \textbf{Supplementary Information}   
\\
\large \textbf{Synthesis and stability of high--$T_c$ LaH$_{10\pm\delta}$ films at high pressures}
\end{center}


\renewcommand{\thefigure}{S\arabic{figure}}
\renewcommand{\thetable}{S\arabic{table}}
\renewcommand{\thesection}{S \Roman{section}}
\renewcommand{\theequation}{S\arabic{equation}}
\makeatletter
\renewcommand{\theHfigure}{S\arabic{figure}}
\renewcommand{\theHtable}{S\arabic{table}}
\renewcommand{\theHsection}{S\arabic{section}}
\renewcommand{\theHequation}{S\arabic{equation}}
\makeatother
\setcounter{figure}{0}
\setcounter{table}{0}
\setcounter{section}{0}
\setcounter{equation}{0}

\begin{figure}[ht!]
    \centering
    \includegraphics[width=\textwidth]{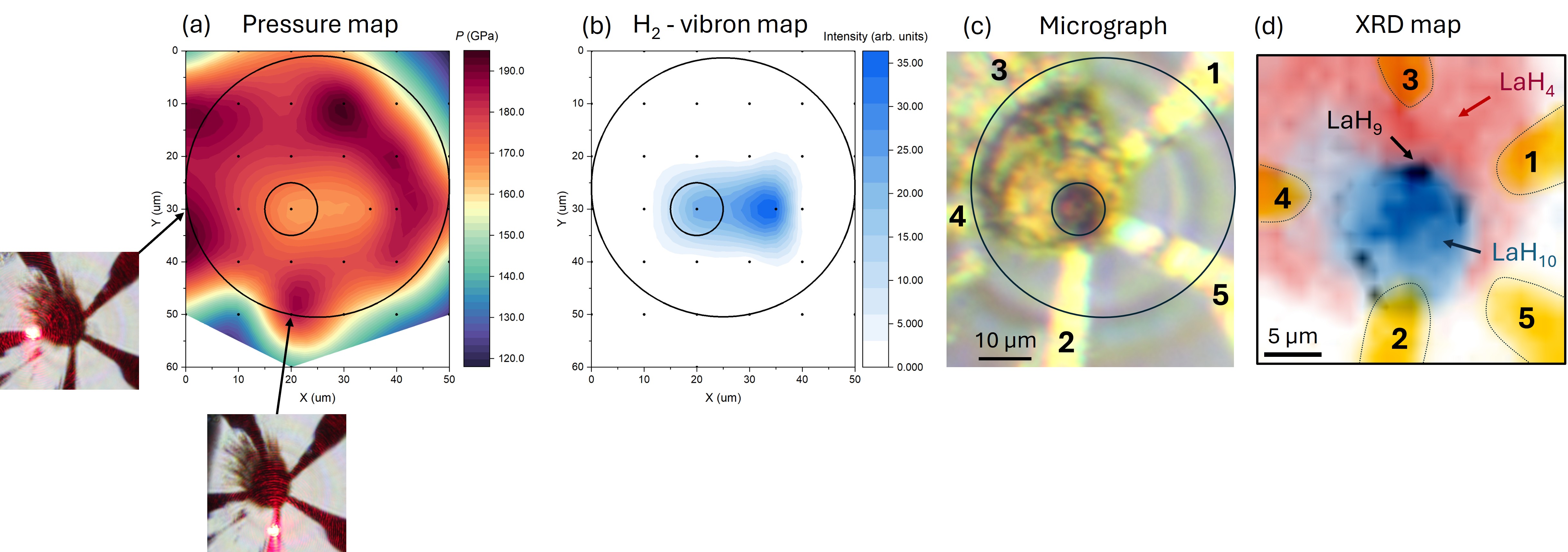}
    \caption{(a) Raman pressure map for DAC 1, with pressure estimated at various positions over the culet using the Akahama scale \cite{akahama_pressure_2006}. Larger outer circle marks the approximate culet edge, while the inner smaller circle marks the dark region observed on the sample after laser heating. (b) Raman map of the molecular hydrogen vibron intensity in DAC 1. Pressure and vibron mapping collected 146 days after laser heating. (c) Photomicrograph of the sample in DAC 1 in the same orientation and on the same scale as in (a) and (b). (d) X-ray diffraction mapping of the sample in DAC 1, with the three identified hydride phases labelled.
    }
    \label{fig: DAC 1_Pressure_Vibron}
\end{figure}

\begin{figure}[ht!]
    \centering
    \includegraphics[width=0.8\textwidth]{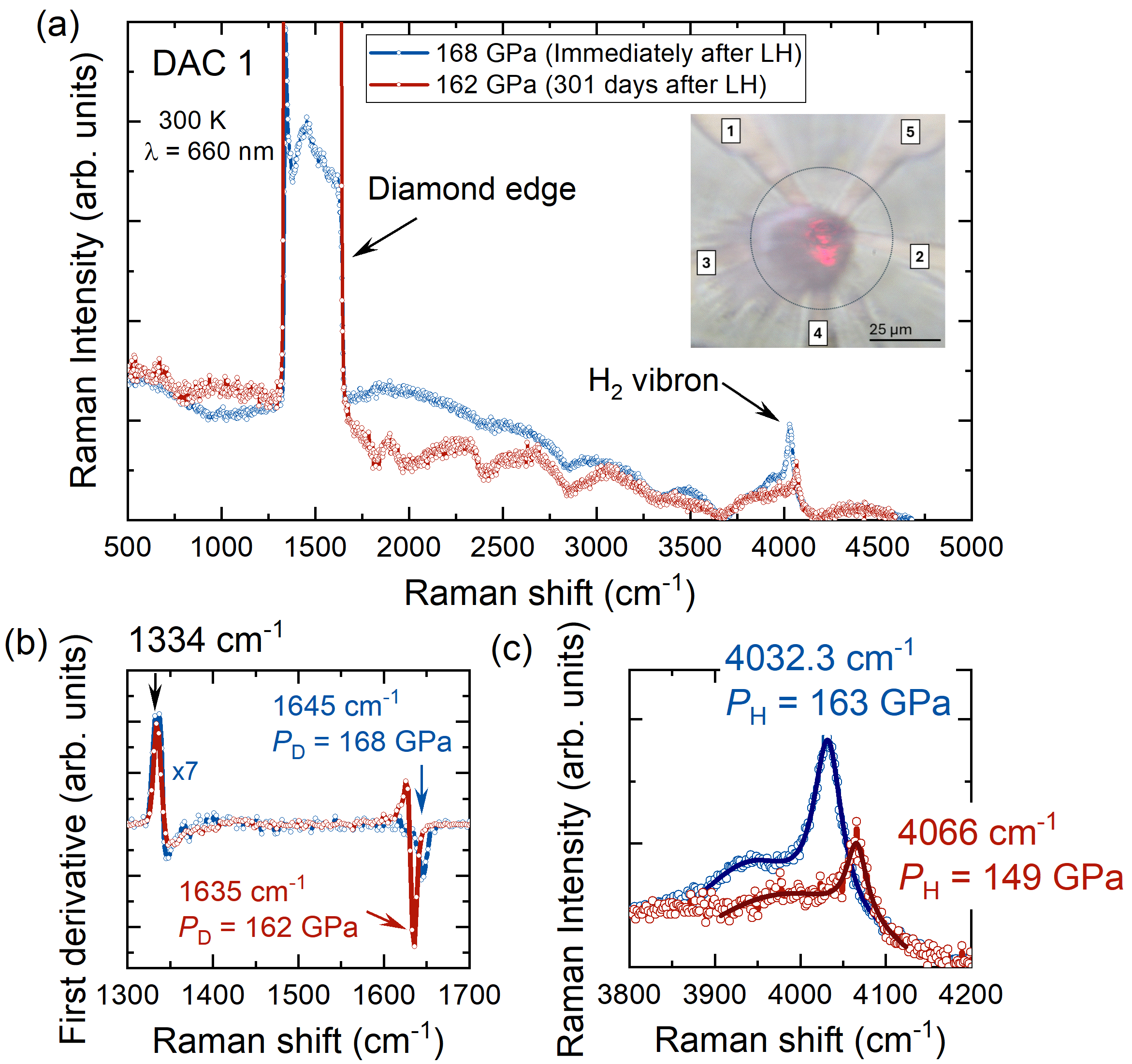}
    \caption{(a) Raman spectra for DAC 1 showing the first order diamond edge and hydrogen vibron immediately after laser heating (blue curve) and after 301 days post laser heating (red curve). Inset shows the sample and Raman laser spot ($\lambda=\SI{660}{\nano\metre}$). (b) First derivative minima used to measure the pressure according to the Akahama scale \cite{akahama_pressure_2006}. (c) Variation in the wavenumber of the hydrogen vibron over the measurement period. Estimates of the pressure from the hydrogen vibron, $P_H$, were made according to Ref. \cite{eremets_universal_2023}.
    }
    \label{fig:DAC 1_Raman stability}
\end{figure}

\begin{figure}[ht!]
    \centering
    \includegraphics[width=0.8\textwidth]{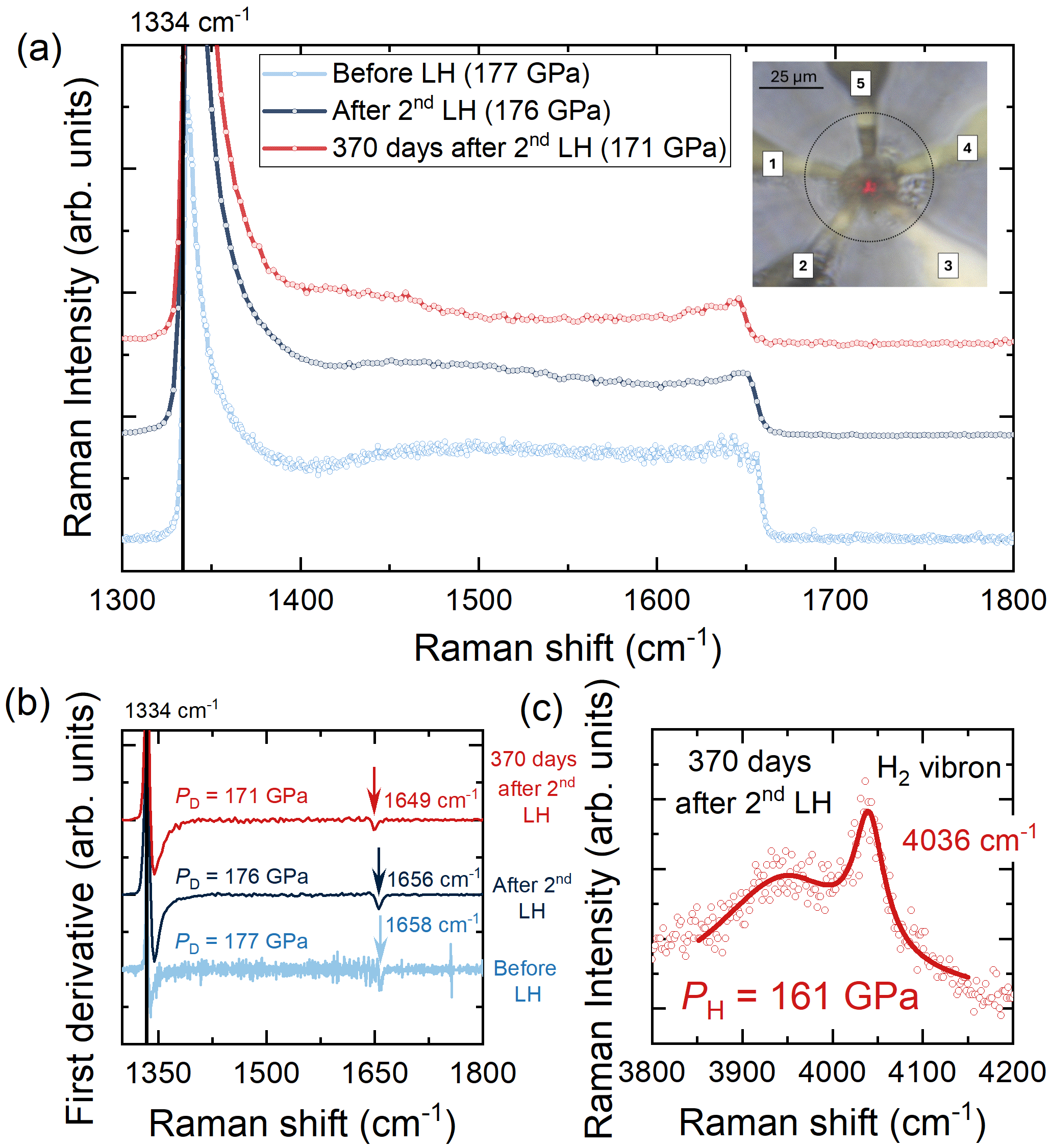}
    \caption{Raman spectra of DAC 2 showing the diamond edge before laser heating, immediately after the second laser heating, and 370 days after the second laser heating. Curves have been offset for clarity. (b) The derivative of the spectra in (a) to determine the pressure from the derivative minima according to the Akahama calibration \cite{akahama_pressure_2006}. The pressure decreased by about \SI{5}{\GPa} after 370 days post laser heating. (c) Evidence of the hydrogen vibron, confirming the long-term retention of excess H$_2$ in DAC 2. The pressure estimated from the hydrogen vibron, $P_H$, using Ref. \cite{eremets_universal_2023} is indicated.}
    \label{fig: DAC 2_Pressure Raman}
\end{figure}

\begin{figure}[ht!]
    \centering
    \includegraphics[width=\textwidth]{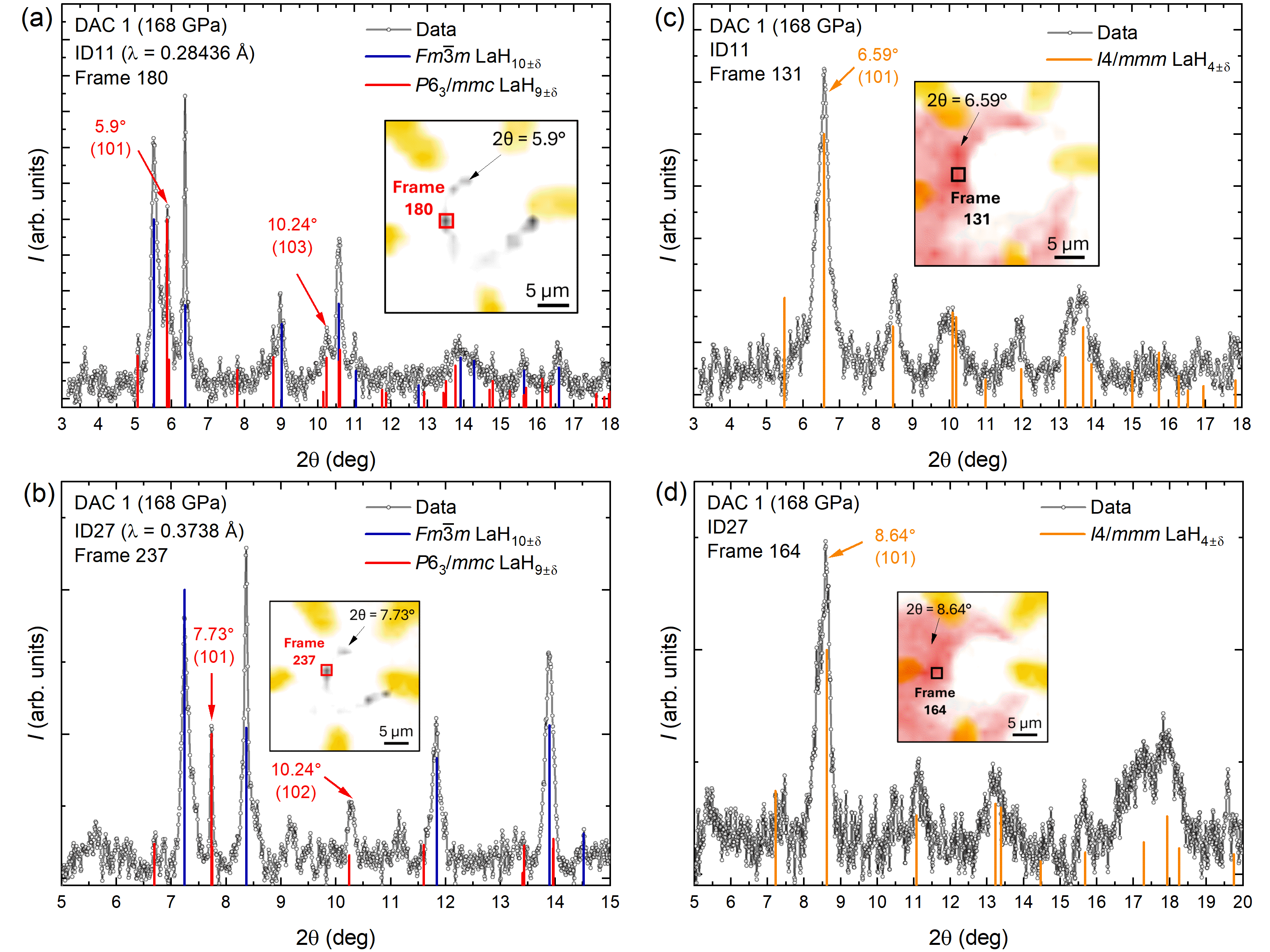}
    \caption{Integrated XRD patterns at different positions on the sample in DAC 1 at \SI{168}{\GPa}, providing evidence for $P6_3/mmc$ and $I4/mmm$ phases in addition to the high-$T_c$ $Fm\bar3m$ phase. (a) and (b) demonstrate the presence of $P6_3/mmc$- LaH$_{9\pm\delta}$ from measurements on ESRF beamlines ID11 (134 days after laser-heating) and ID27 (195 days after laser-heating) respectively. The strong (101) reflection of the $P6_3/mmc$ phase present between the first two $Fm\bar3m$ reflections has been observed in other studies of LaH$_{10}$ \cite{sun_high-temperature_2021, drozdov_superconductivity_2019}. Inset spatial map of the (101) reflection's intensity shows the localised region of the phase, and red squares indicate the frame positions where integrated patterns were measured. From the ID11 measurement, we estimate the unit cell parameters to be $a=b\sim \SI{3.71}{\angstrom},\, c \sim \SI{5.50}{\angstrom},\, V\sim\SI{65.7}{\angstrom\cubed},\, c/a\sim1.48$, but do not perform full structure refinement owing to poor resolution of the phase. (c) and (d) Evidence for the $I4/mmm$-LaH$_{4\pm\delta}$ phase in DAC 1. Inset spatial maps show the distribution of the phase by mapping of the intense (101) reflection. From the ID11 measurements of the $I4/mmm$ phase, we estimate $a=b\sim \SI{2.73}{\angstrom},\, c \sim \SI{5.94}{\angstrom},\, V\sim\SI{44.2}{\angstrom\cubed},\, c/a\sim2.18$.
    }
    \label{fig: DAC 1_representative frames}
\end{figure}

\begin{figure}[ht!]
    \centering
    \includegraphics[width=0.7\textwidth]{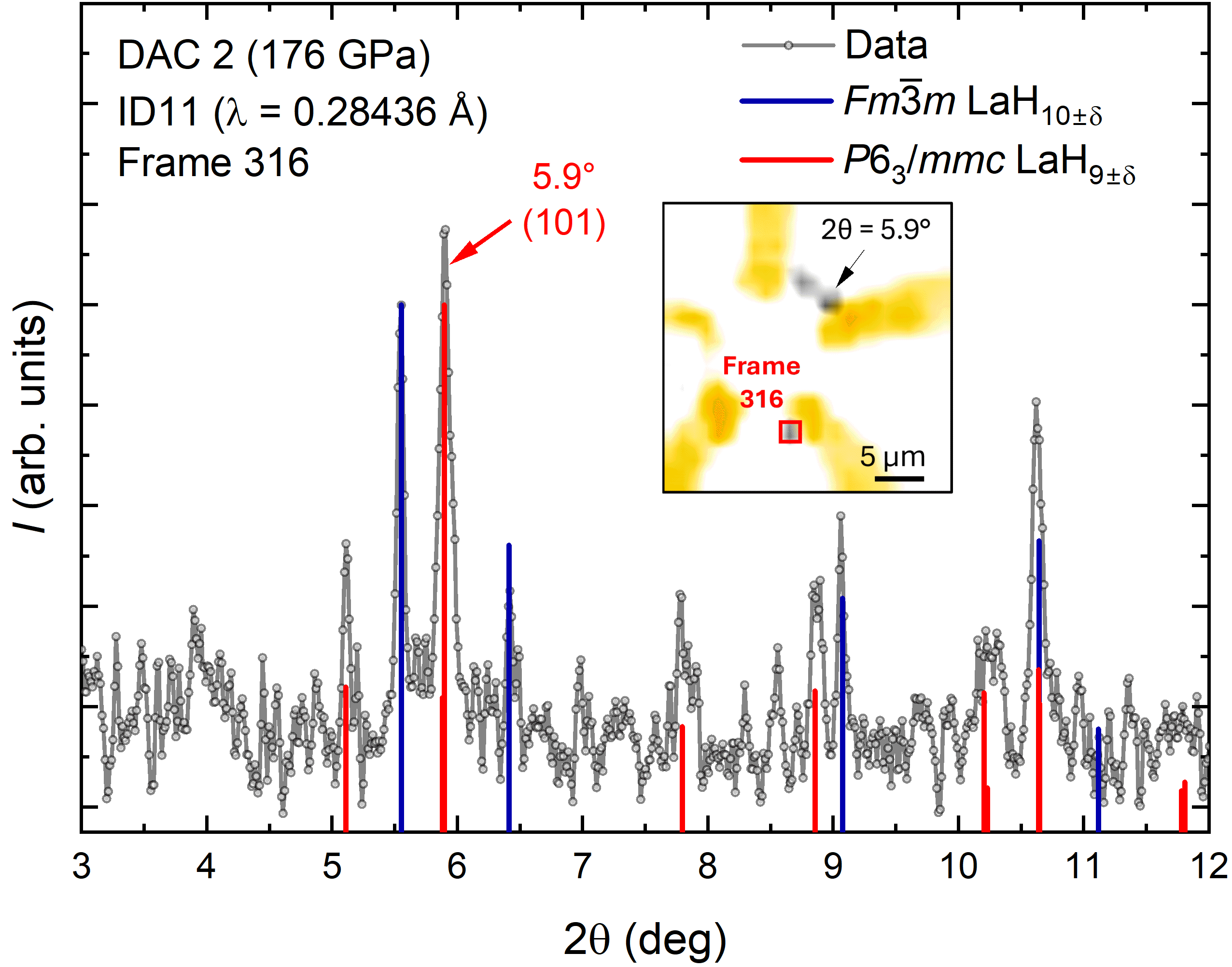}
    \caption{Evidence for the $P6_3/mmc$- LaH$_{9\pm\delta}$ phase in DAC 2 at \SI{176}{\GPa}. Measurements performed 43 days after the second laser heating. Inset spatial maps show the distribution of the phase by mapping of the intense (101) reflection at $2\theta=\SI{5.9}{\degree}$. We estimate $a=b\sim \SI{3.68}{\angstrom},\, c \sim \SI{5.54}{\angstrom},\, V\sim\SI{65.1}{\angstrom\cubed},\, c/a\sim1.51$.
    }
    \label{fig: DAC 2_hcp LaH9}
\end{figure}

\begin{figure}[ht!]
    \centering
    \includegraphics[width=0.7\textwidth]{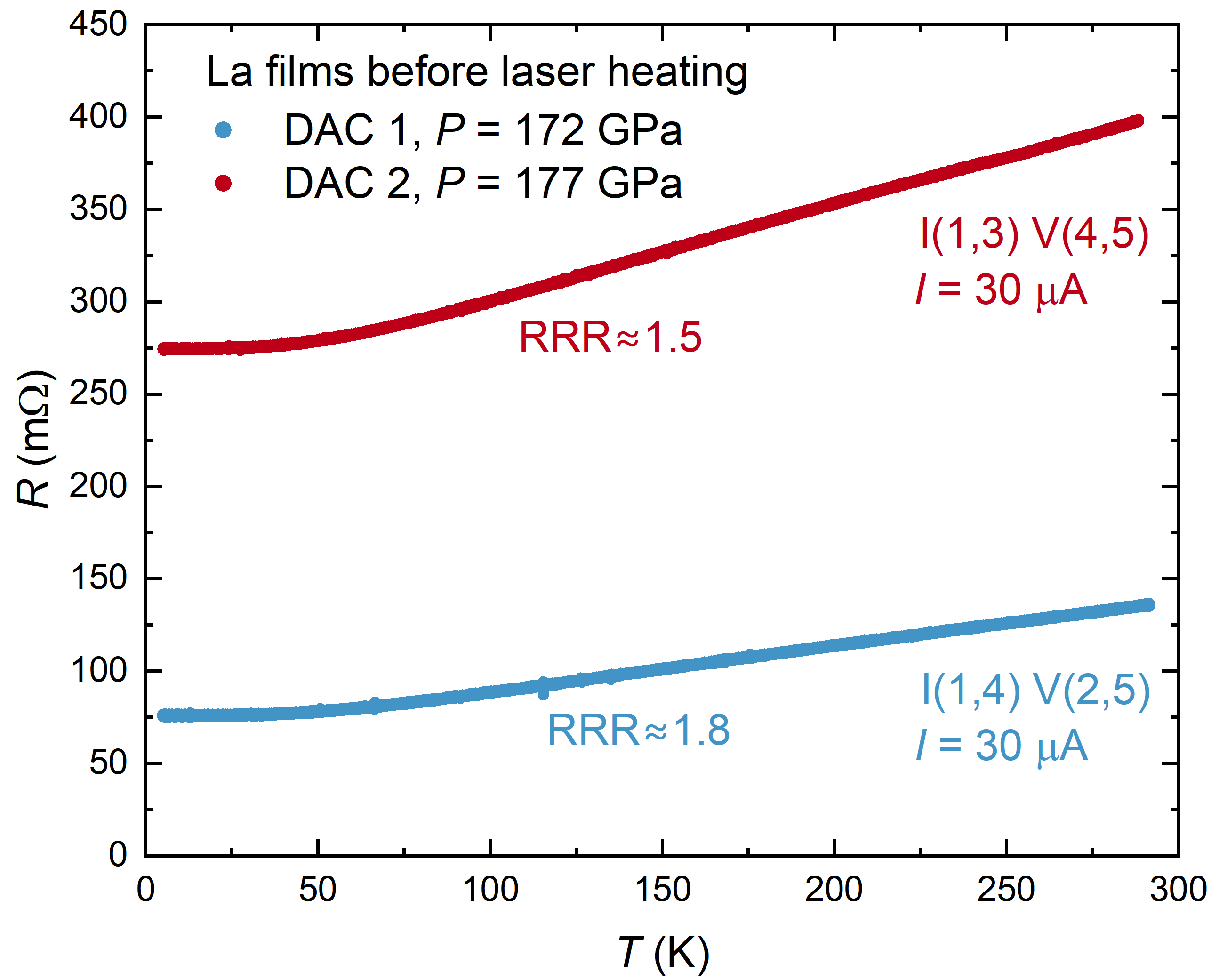}
    \caption{Resistance measurements of the elemental lanthanum films in both DACs prior to laser heating, showing metallic behaviour. Measurements taken with excitation current of \SI{30}{\micro\ampere} in warming runs at rates between 0.5-\SI{1}{\kelvin\per\minute}. The residual resistance ratios (RRR = $R_{\SI{298}{\kelvin}}/R_{\SI{5}{\kelvin}}$), a measure of the quality of the films, are indicated for the respective samples. 
    }
    \label{fig:RTs La before LH_SI}
\end{figure}

\begin{figure}[ht!]
    \centering
    \includegraphics[width=0.7\textwidth]{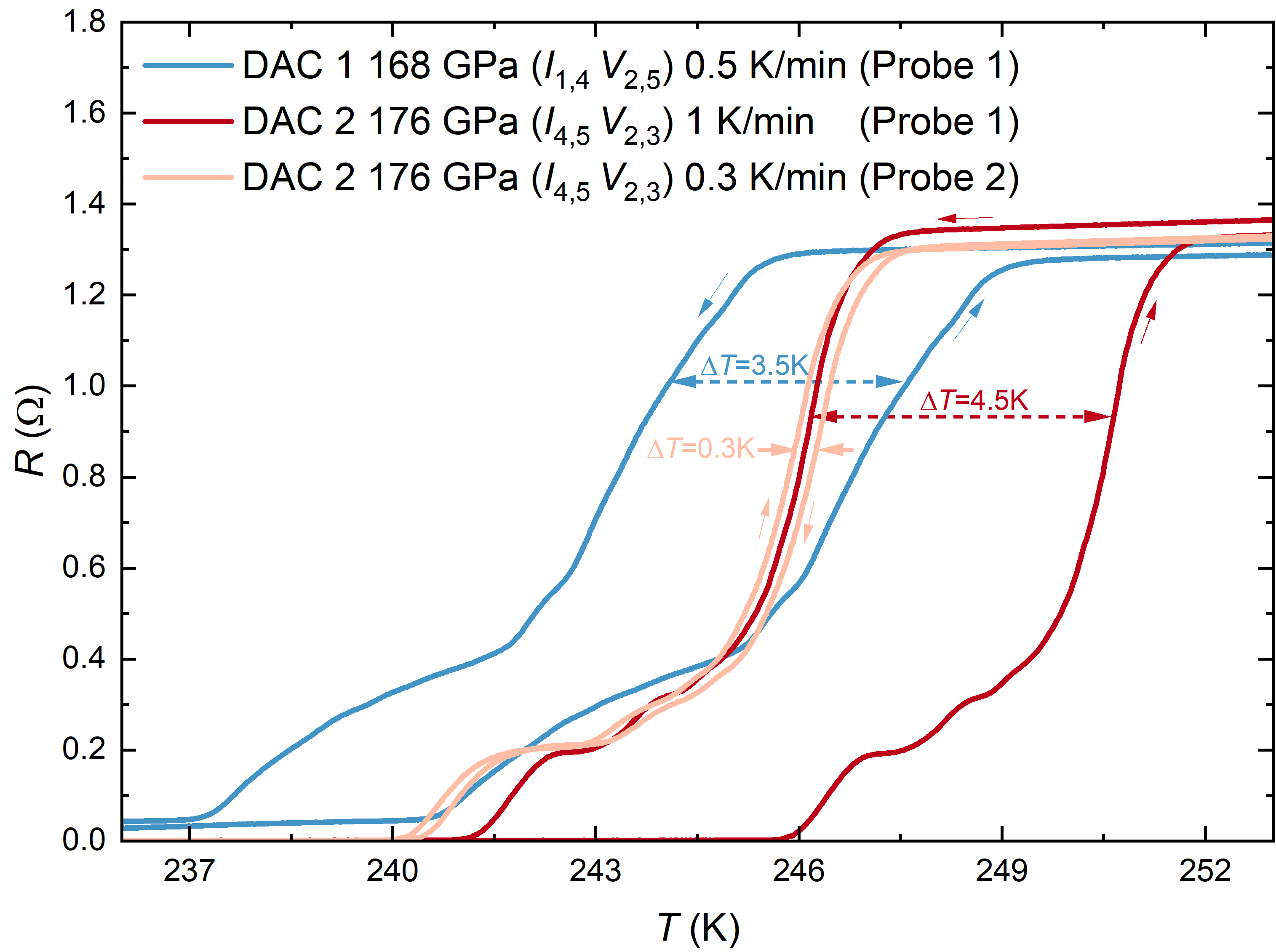}
    \caption{Hysteresis between cooling and warming runs measured at different temperature sweep rates for both DACs, using two different measurement probes (probes 1 and 2). Zero-field measurements were performed using probe 1, where the DAC is surrounded by He gas at \SI{1}{\bar}. A Cernox thermometer was mounted on the probe cold plate, to which the DAC is attached. A \SI{3.5}{\kelvin}--\SI{4.5}{\kelvin} temperature hysteresis in $T_c$ is observed with this probe. Measurements in magnetic field were performed using probe 2, which is custom designed for a magnet cryostat, with the sample under vacuum. In this case, a Cernox thermometer was mounted directly on the DAC seat at the base of the diamond directly next to the sample, resulting in a much smaller temperature hysteresis between cooling and warming. Owing to the improved thermal coupling between the thermometer and the sample, the $T_c$ measured with probe 2 is expected to provide the most accurate estimate of the intrinsic superconducting transition temperature. This value is in excellent agreement with the $T_c$ measured during cooling using probe 1. For this reason, all measurements acquired using probe 1 and presented in the main text are shown for the cooling runs only. Arrows on each curve indicate the cooling and warming directions.  
    }
    \label{fig:RT_hysteresis}
\end{figure}

\begin{figure}[ht!]
    \centering
    \includegraphics[width=\textwidth]{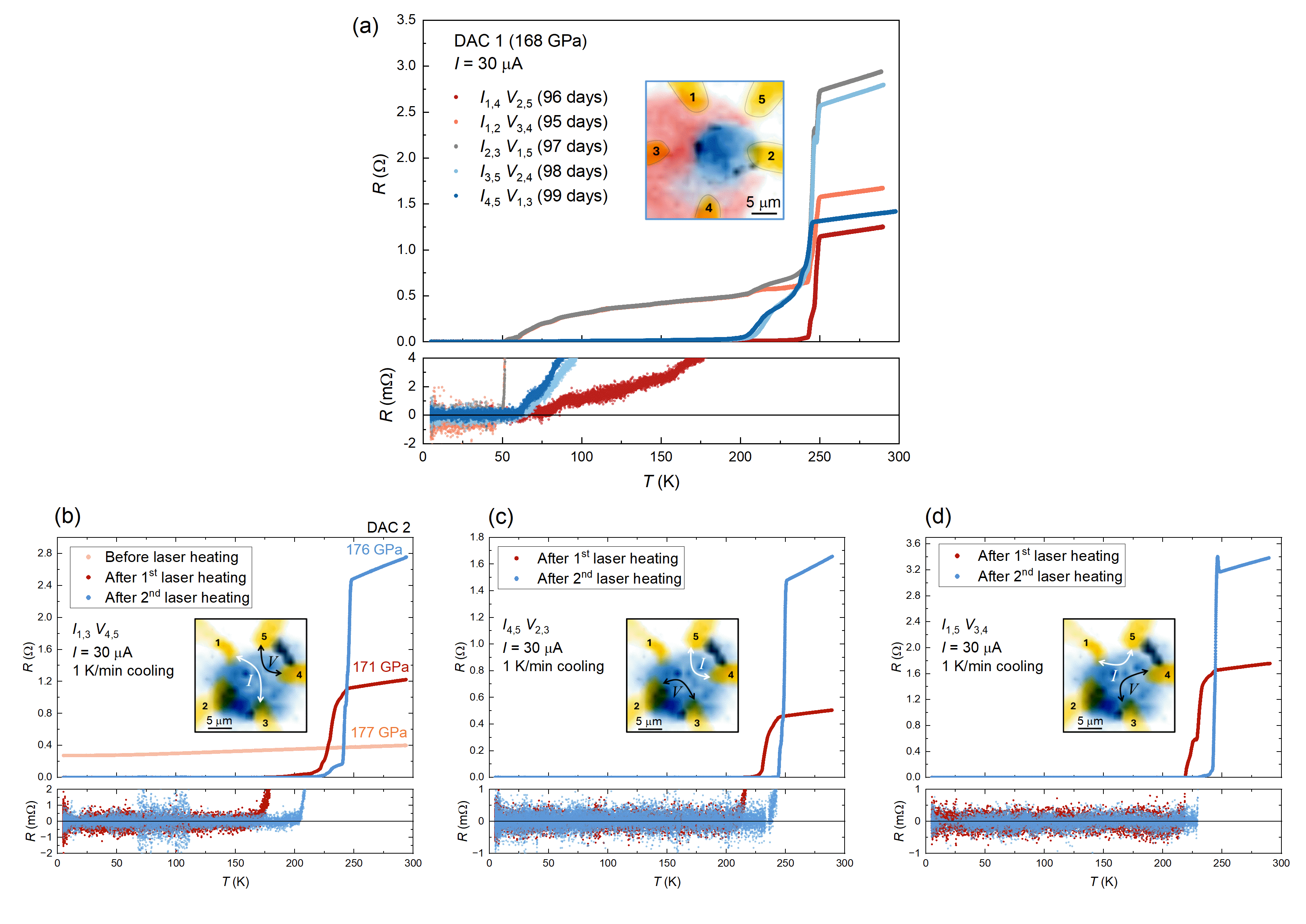}
    \caption{Resistive transitions in various electrode configurations. (a) Measurements for DAC 1, where the current/voltage contacts and days since laser heating are indicated in the legend. The inset spatial XRD map illustrates the contact numbering relative to the observed phases, where the blue region is $Fm\bar3m$-LaH$_{10\pm\delta}$, red is $I4/mmm$-LaH$_{4\pm\delta}$ and black is $P6_3/mmc$-LaH$_{9\pm\delta}$. (b)--(d) Measurements for DAC 2 after first and second laser heating cycles in three different measurement configurations. The electrode configuration used for each measurement is shown on the XRD spatial map in the inset of each plot. The pressure after the first and second laser heating of DAC 2 was \SI{171}{\GPa} and \SI{176}{\GPa} respectively. Successive laser heating cycles caused the normal state resistance to increase, and the resistive transition to sharpen, with an onset $T_c$ at \SI{247}{\kelvin} after the second laser heating. Zero resistance within the noise floor is observed in all measured configurations of DAC 2, consistent with the $Fm\bar3m$ phase forming a continuous pathway between all electrodes.
    }
    \label{fig: Config summary}
\end{figure}

\begin{figure}[ht!]
    \centering
    \includegraphics[width=\textwidth]{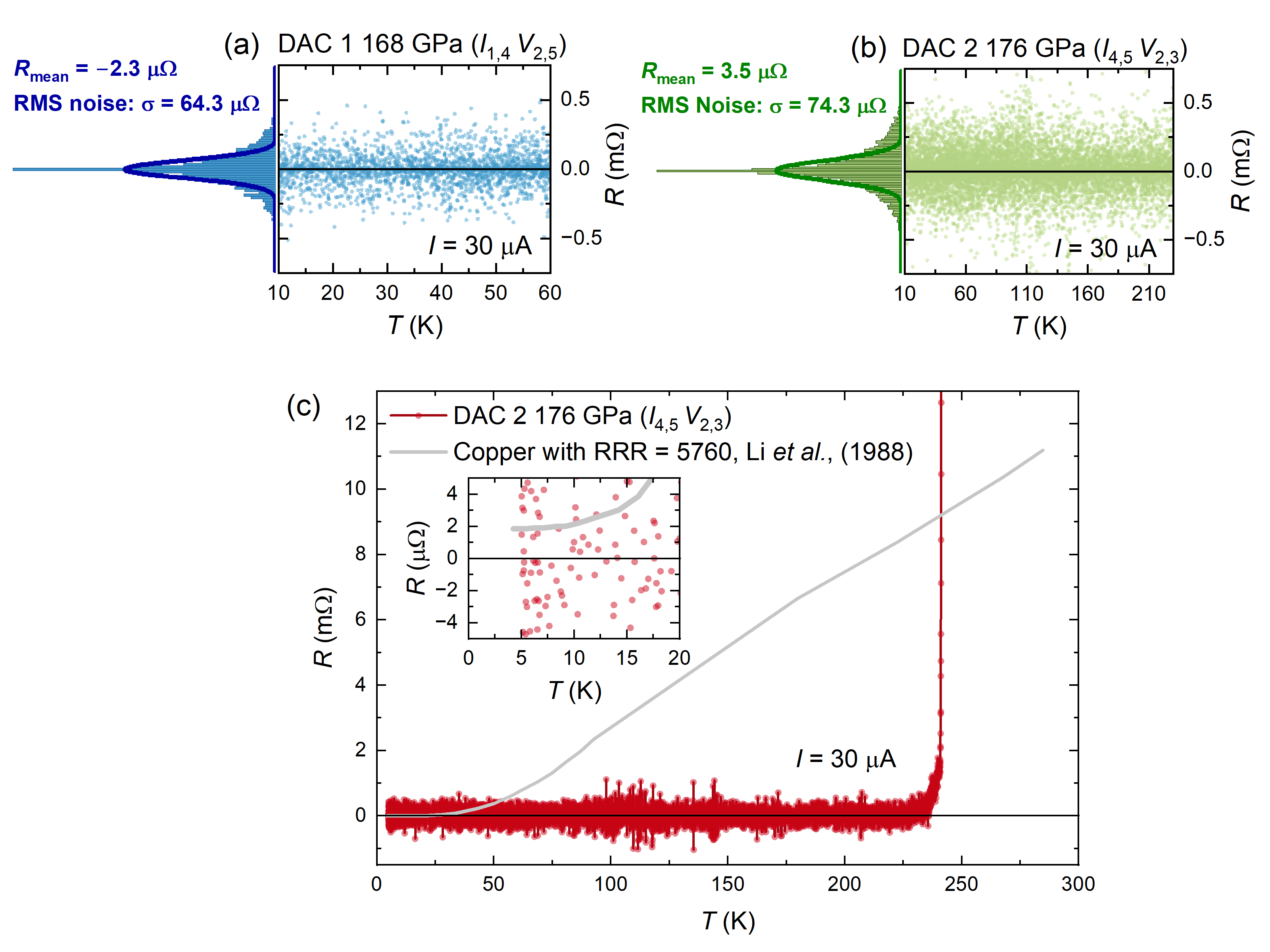}
    \caption{Quantitative confirmation of zero resistance in LaH$_{10\pm\delta}$. 
(a) and (b) Resistance curves as a function of temperature with the corresponding histograms of resistance residuals in the zero-resistance state for DAC~1 and DAC~2. 
The RMS noise is extracted as $RMS\sim\sigma$, where $\sigma$ is the standard deviation obtained from a Gaussian fit to the distribution. 
The measured RMS noise is consistent with the input noise of the SRS SIM921 preamplifier (JFET LSK489), which we calculate to be \SI{60}{\micro\ohm} with \SI{30}{\micro\ampere} excitation and a \SI{1}{\second} time constant. 
The non-Gaussian tails observed in both distributions are not expected from instrumental noise and indicate additional contributions from the measurement environment. 
(c) Comparison of the resistance of DAC~2 and high-purity copper as a function of temperature. 
The copper resistance was estimated using the resistivity $\rho$ from Ref.~\cite{li_thermal_1988} and the measured dimensions of the LaH$_{10\pm\delta}$ region in DAC 2 ($l\sim\SI{3.9}{\micro\metre}$, $w\sim\SI{26}{\micro\metre}$, and $t=\SI{213.5}{\nano\metre}$) according to $R=\rho l/wt$. 
The residual resistivity ratio of the high-purity copper sample is $RRR=\rho_{\SI{293}{\kelvin}}/\rho_{\SI{4.2}{\kelvin}}=5760$.
}
    \label{fig: Noise analysis_all DACs}
\end{figure}

\begin{figure}[ht!]
    \centering
    \includegraphics[width=\textwidth]{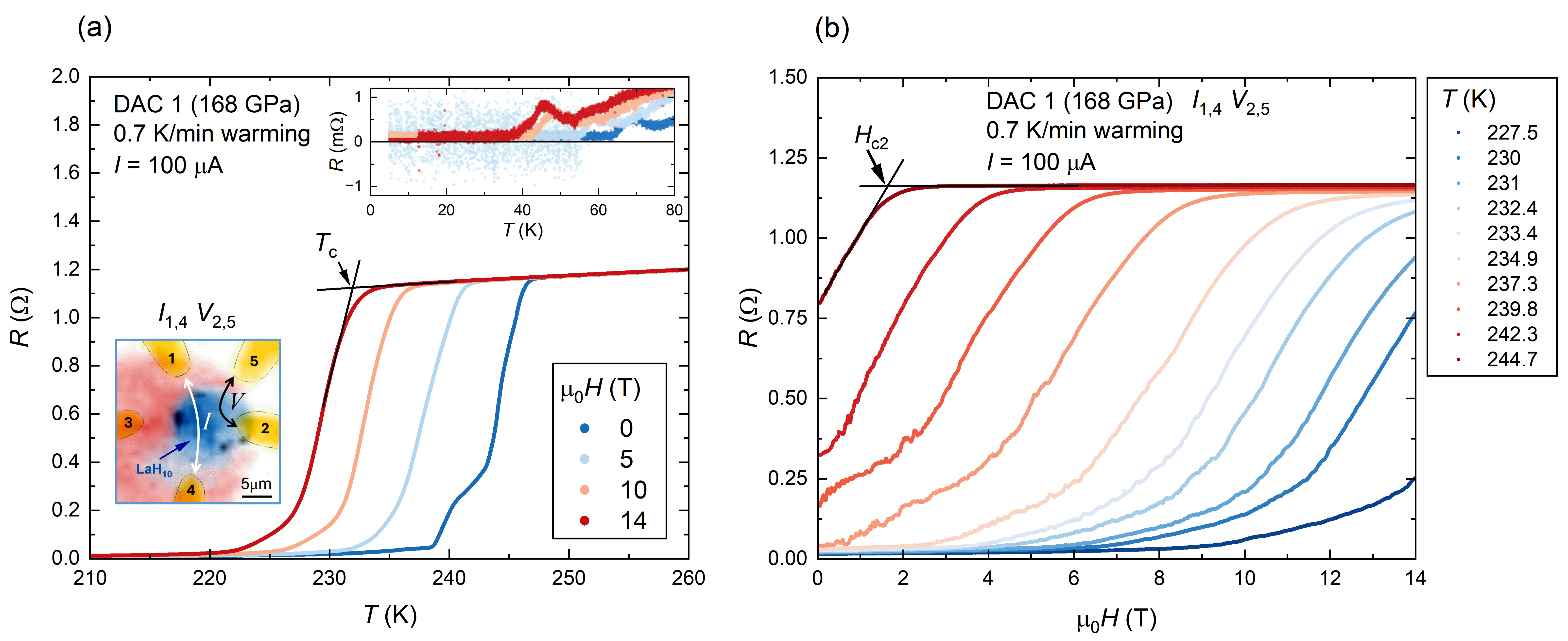}
    \caption{Suppression of the superconducting transition of the LaH$_{10\pm\delta }$ film in DAC 1 at \SI{168}{\giga\pascal}, measured using probe 2 (see Fig. \ref{fig:RT_hysteresis} caption). (a) Temperature sweeps in static magnetic fields up to \SI{14}{\tesla}, demonstrating the downward shift in critical temperature $T_c$. The value of $T_c$ was extracted as the intersection between linear fits to the normal state and superconducting transition as indicated for the \SI{14}{\tesla} curve. The XRD mapping of the sample is shown in the inset with the electrode configuration used for the measurements. Top inset plot shows the low temperature regime. Saturation of the resistance signal and variations in the noise level were caused by a sensitivity auto-ranging issue in the data acquisition program. At temperatures below \SI{60}{\kelvin} in zero field, we observe a positive offset of approximately \SI{125}{\micro\ohm} in the resistance noise floor. We find that this offset is consistent with common-mode voltage arising from the finite resistance of the current leads, which is approximately \SI{20}{\ohm} per lead. At the applied excitation current of \SI{100}{\micro\ampere}, this offset corresponds to a common-mode voltage of \SI{12.5}{\nano\volt}.
    The Stanford Research Systems Low-Noise Voltage Preamplifier SR560 has a common-mode rejection ratio (CMRR) of \SI{100}{\deci\bel} up to \SI{1}{\kilo\hertz}. Using this specification, we estimate that a common-mode voltage of \SI{12.5}{\nano\volt} would arise from a common-mode source corresponding to a resistance of approximately \SI{12.5}{\ohm}, which is of the same order as the measured resistance of our current leads. We therefore conclude that the finite resistance offset does not originate from the sample, but from imperfect rejection of the common-mode voltage associated with the current leads.
    (b) Isothermal field sweeps for the sample in DAC 1. The value of $H_{c2}$ was extracted as indicated for the curve measured at $T = \SI{244.7}{\kelvin}$.
    }
    \label{fig: Hc2_DAC1_SI}
\end{figure}

\begin{figure}[ht!]
    \centering
    \includegraphics[width=\textwidth]{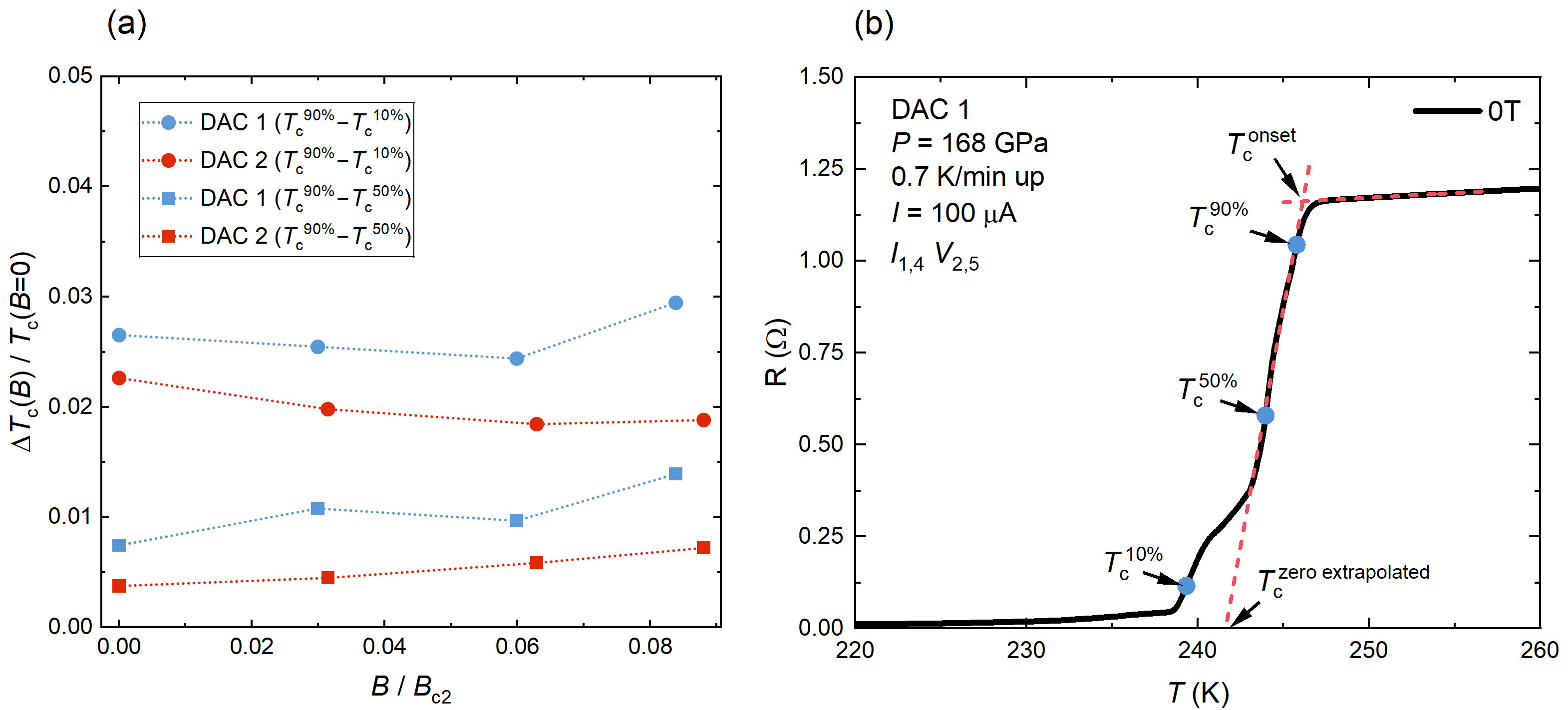}
    \caption{Broadening of the superconducting transition under external magnetic fields. 
    We calculate $\frac{\Delta T_c(B)}{T_c(B=0)}$ as defined in Ref. \cite{talantsev2024comment} using a 90\% and 10\% criteria: $\frac{T_c^{90\%}(B)-T_c^{10\%}(B)}{\big[T_c^{90\%}(0)+T_c^{10\%}(0)\big]/2}$, and also using a 90\% and 50\% criteria for comparison: $\frac{T_c^{90\%}(B)-T_c^{50\%}(B)}{\big[T_c^{90\%}(0)+T_c^{50\%}(0)\big]/2}$. Using the 90\% and 10\% criteria, we find that the transition widths for both DAC 1 and 2 slightly decrease in fields up to \SI{10}{\tesla} and begin to broaden at \SI{14}{\tesla}. This behaviour is similar to behaviour reported in Ref. \cite{talantsev2024comment}. However, using the 90\% and 50\% criteria, the transition widths increase monotonically in magnetic field. The maximum applied magnetic field here is 14T which is less than 10\% of $B_{c2}$ for DAC 1 and 2. (b) Resistance curve for DAC 1 in zero field, illustrating how the different values of $T_c$ were determined.}
    \label{fig: Tc Broadening in field}
\end{figure}

\begin{figure}[ht!]
    \centering
    \includegraphics[width=0.5\textwidth]{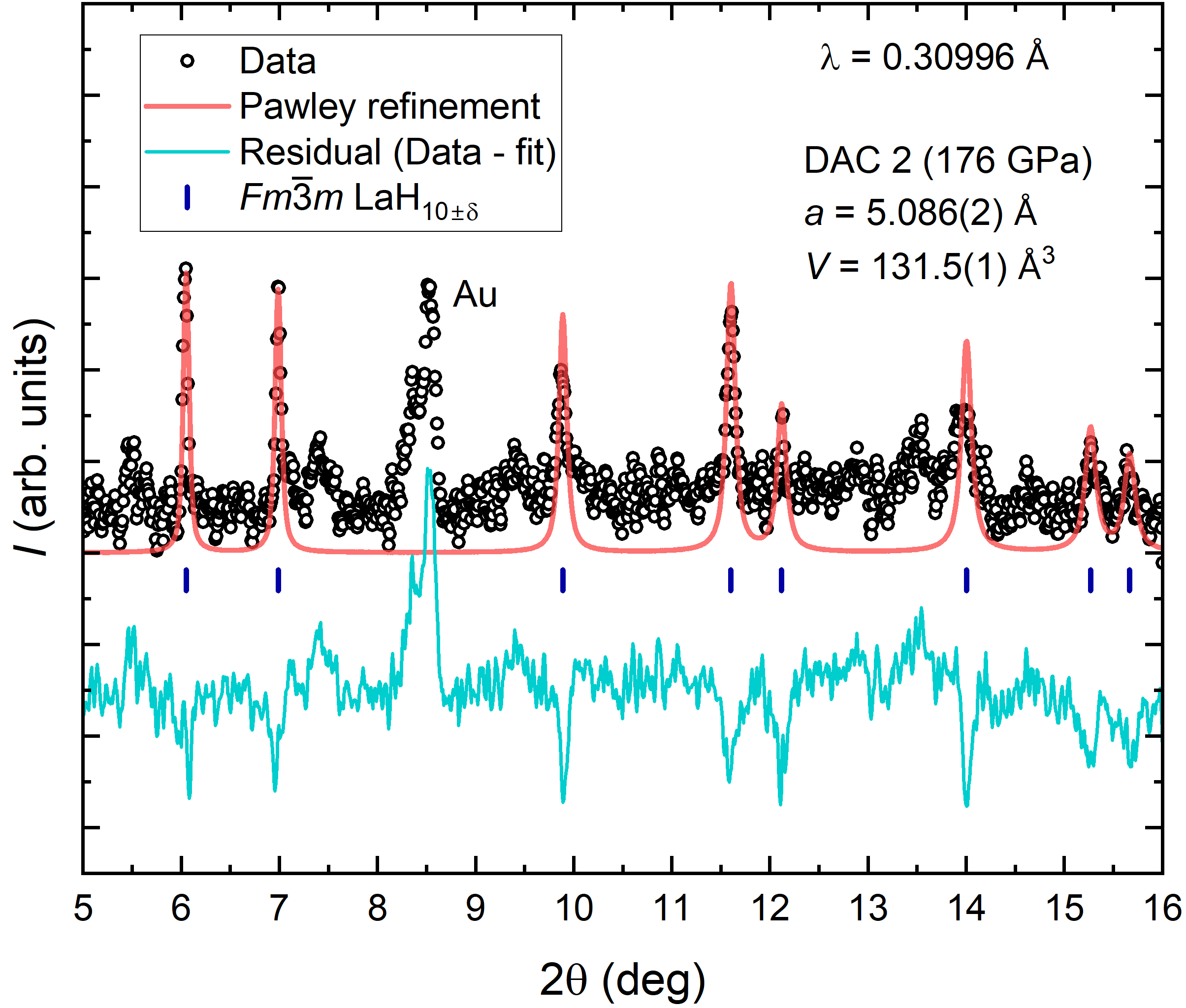}
    \caption{Measurements of DAC 2 on beamline I15 at Diamond Light Source 32 days after the second laser heating. The measured data are an average of five frames over the sample region to improve statistics due to poor resolution of the $Fm\bar3m$ phase in the measurements and the reduced signal to noise from the I15 measurements compared to the ESRF measurements.
    }
    \label{fig: DAC2_June DLS}
\end{figure}

\begin{figure}[ht!]
    \centering
    \includegraphics[width=\textwidth]{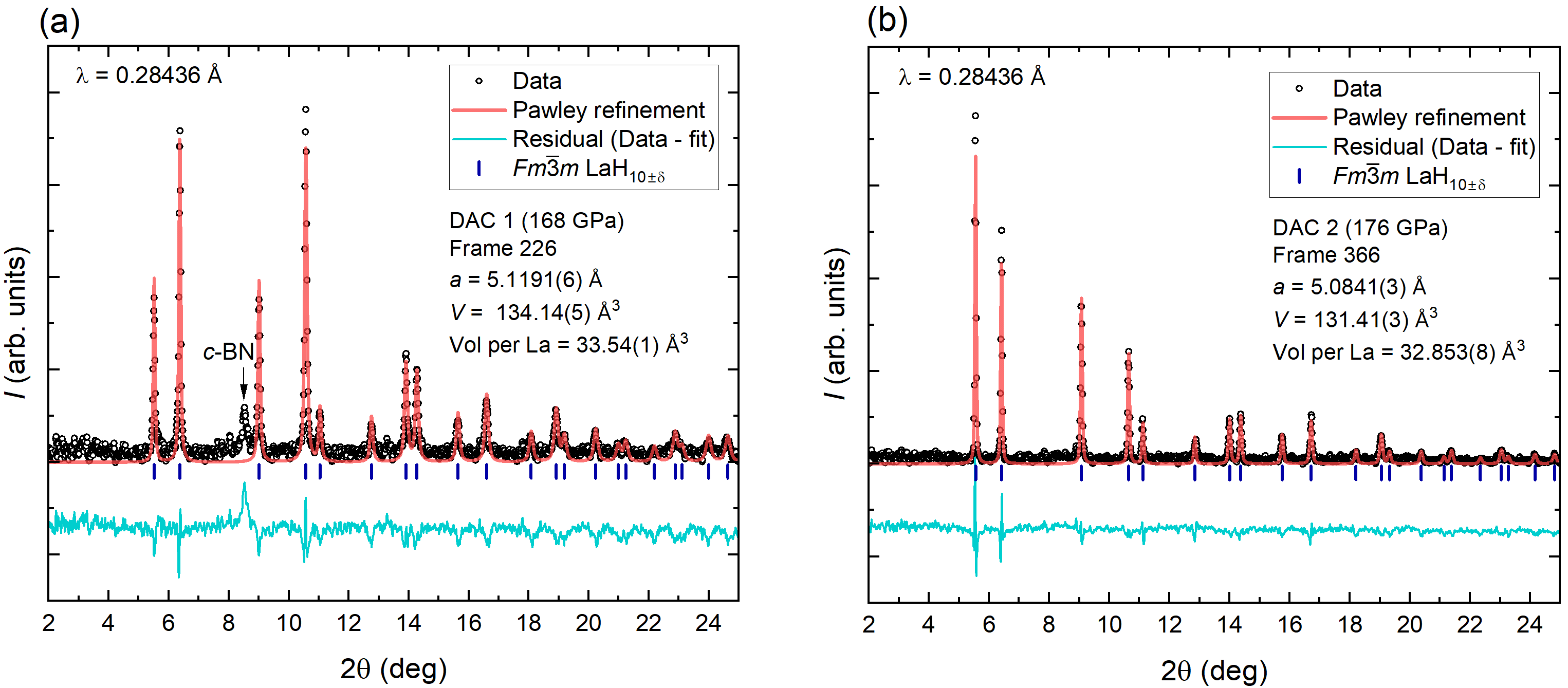}
    \caption{X-ray diffraction measurements performed on beamline ID11 at the ESRF. Pawley refinements of integrated patterns for (a) DAC 1 134 days after laser heating and (b) DAC 2 43 days after laser heating, showing reflections corresponding to $Fm\bar3m$-LaH$_{10\pm\delta}$. The positions on the sample at which the integrated frames were acquired are shown on the XRD mappings in Figs. \ref{fig: ESRF_Mapping comparison}(a) and (c).}
    
    \label{fig: ID11_Summary}
\end{figure}

\begin{figure}[ht!]
    \centering
    \includegraphics[width=\textwidth]{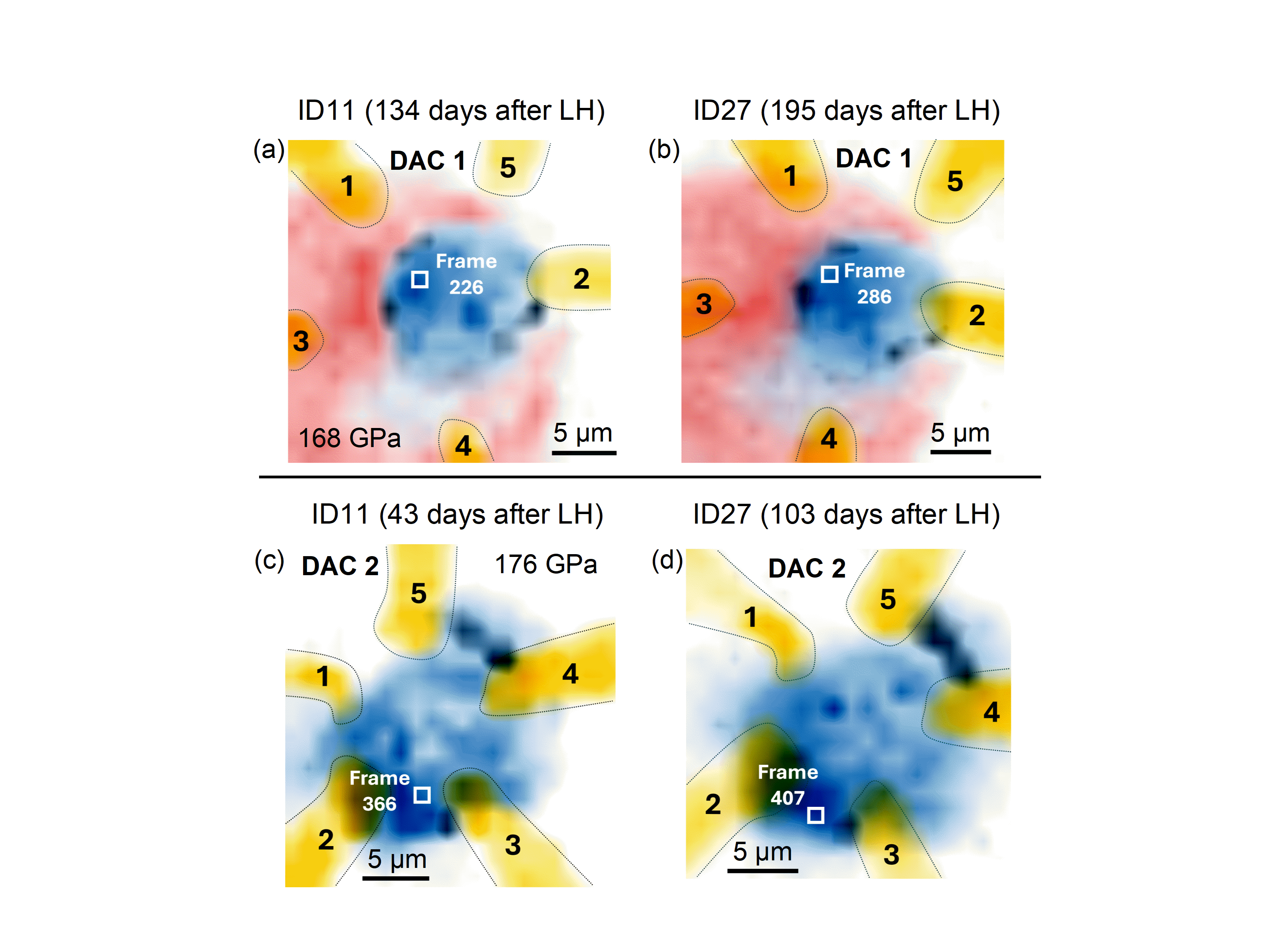}
    \caption{Comparison between high-resolution XRD spatial mappings from ESRF beamlines ID11 and ID27 for DACs 1 (a)-(b), and DAC 2 (c)-(d). Measurements were acquired about 60 days apart. Blue regions correspond to the (311) reflection of $Fm\bar3m-$LaH$_{10\pm\delta}$, localised black regions correspond to the (101) reflection of $P6_3/mmc-$LaH$_{9\pm\delta}$, and the red regions for DAC 1 correspond to the (101) reflection of $I4/mmm-$LaH$_{4\pm\delta}$. No traces of the $I4/mmm$ phase were identified in DAC 2. The spatial map of the electrodes corresponds to the (111) reflection of Au. For each phase, the (hkl) were chosen to ensure no overlap between the identified phases. Frame numbers shown on the spatial maps indicate positions where integrated patterns were measured to refine the unit cell parameters of the $Fm\bar3m$ phase. Visualisation of the spatial maps were performed using XDI \cite{hrubiak_multimode_2019}.
    }
    \label{fig: ESRF_Mapping comparison}
\end{figure}

\begin{figure}[ht!]
    \centering
    \includegraphics[width=0.5\textwidth]{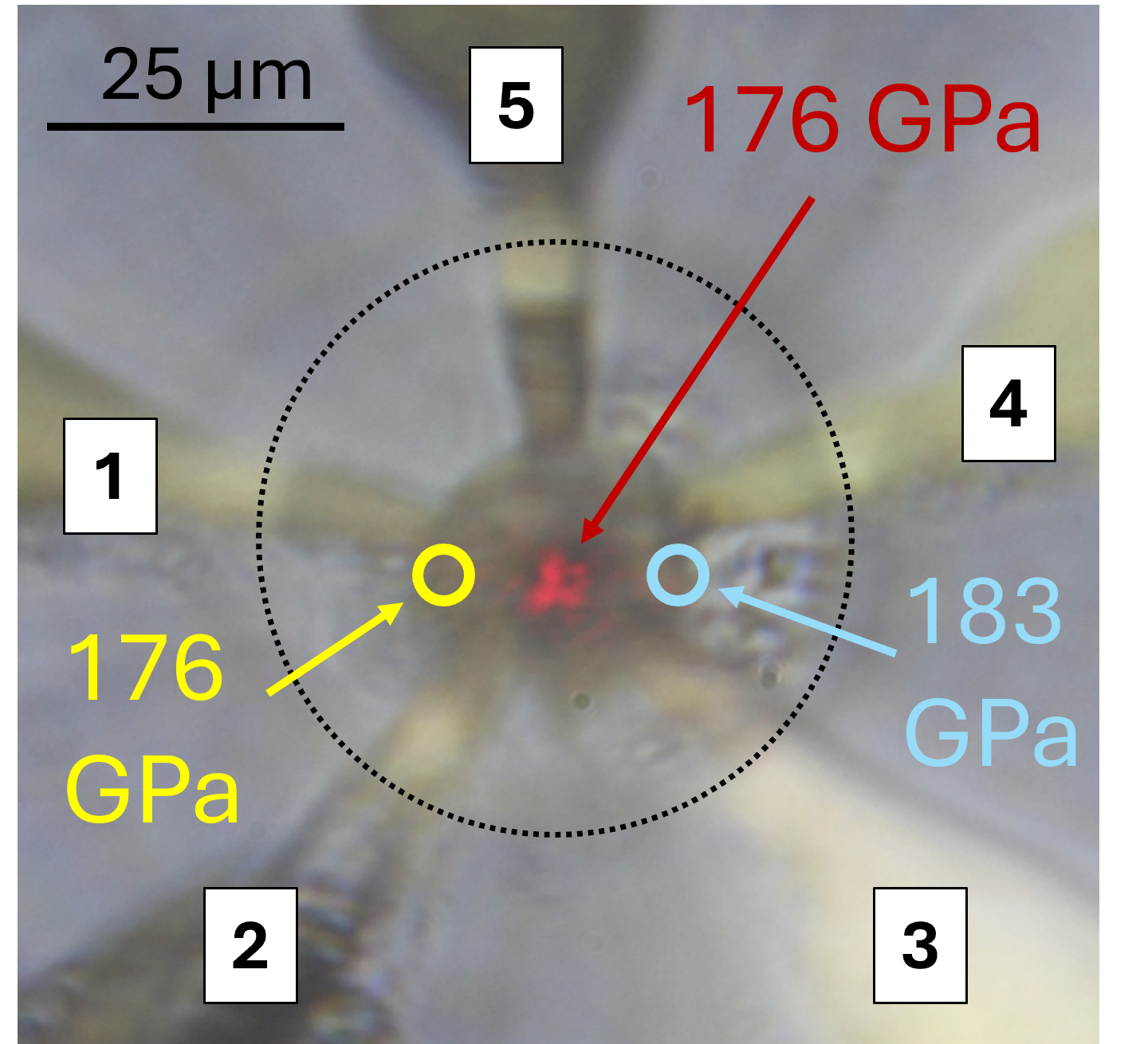}
    \caption{Measurement of pressure at different positions over the sample in DAC 2 using the position of the diamond edge in Raman spectroscopy. Measurements shown are after the second laser heating of DAC 2. Raman laser spot ($\lambda=\SI{660}{\nano\metre}$) is visible at the sample centre. Dotted circle indicates the approximate culet position.
    }
    \label{fig: DAC2_Pressure gradient}
\end{figure}

\begin{figure}[ht!]
    \centering
    \includegraphics[width=0.8\textwidth]{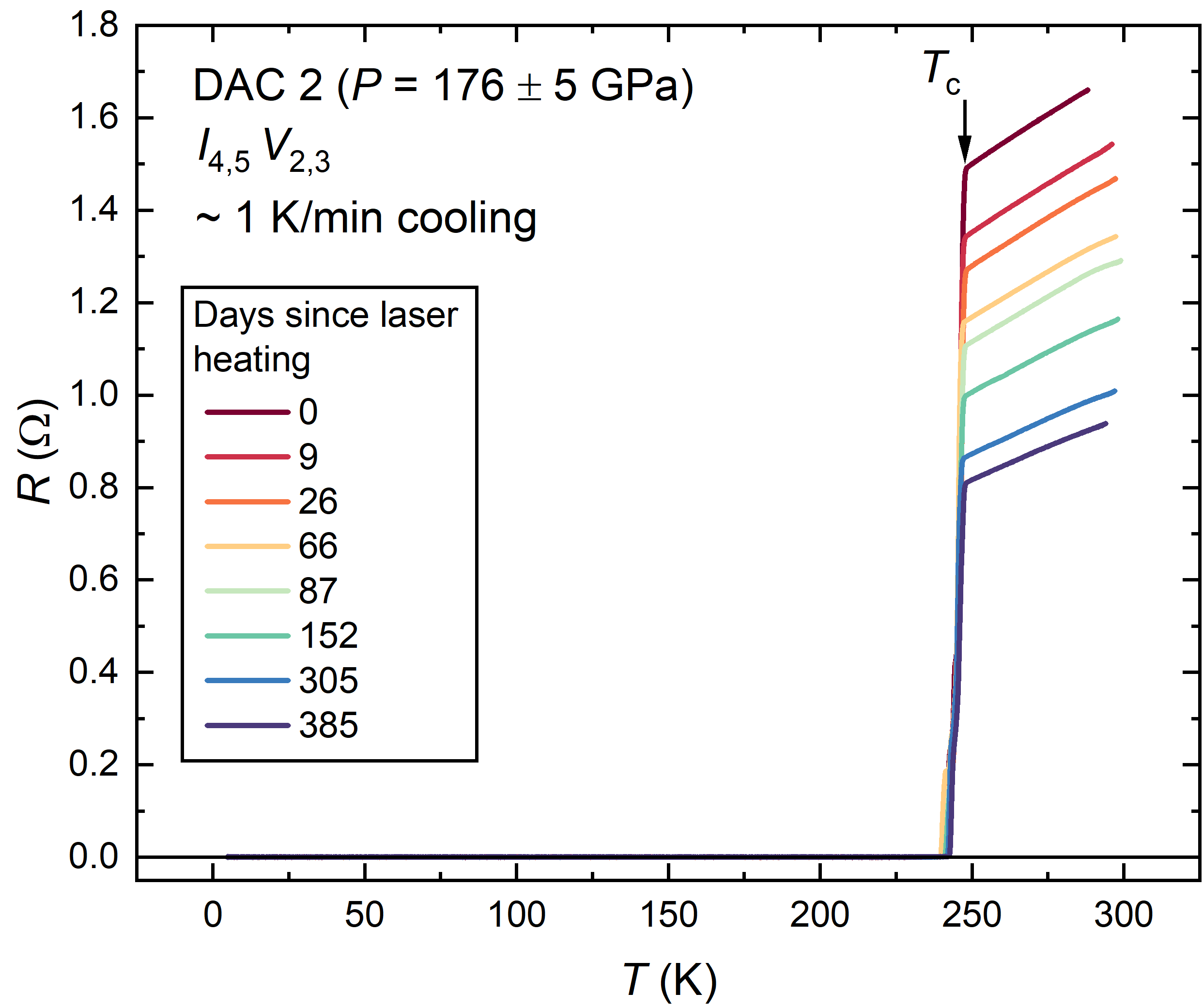}
    \caption{Resistance versus temperature curves for the sample in DAC 2, performed over a period up to 385 days after the second laser heating. $T_c$ remains very stable, while the magnitude of the normal state resistance decreases over time. This could be due to annealing at room temperature, or related to the \SI{5}{\GPa} pressure decrease in the DAC over time.}
    \label{fig: DAC 2_RTs over time.png}
\end{figure}

\begin{figure}[ht!]
    \centering
    \includegraphics[width=0.8\textwidth]{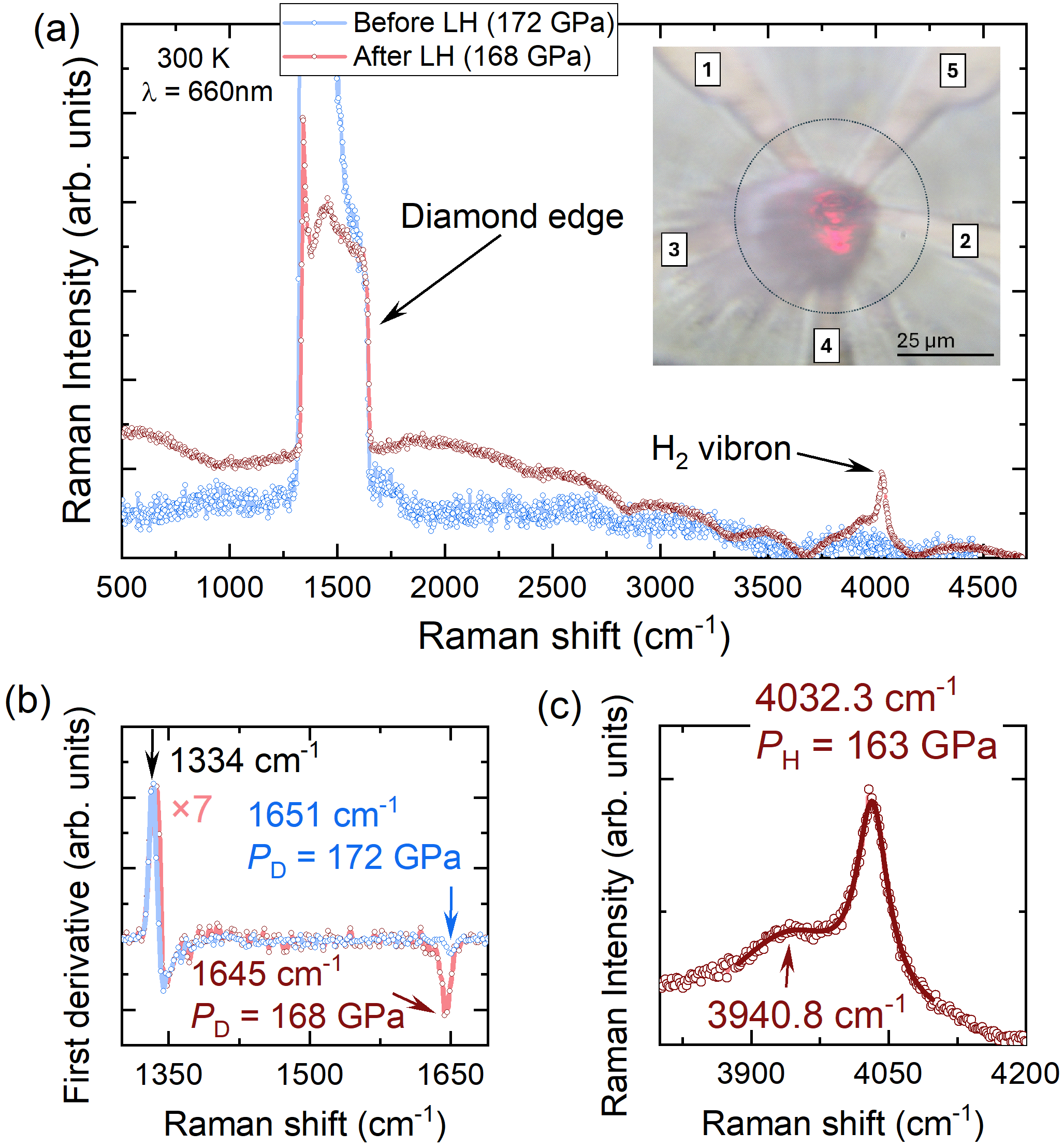}
    \caption{(a) Raman spectrum of DAC 1 before and immediately after laser heating. The edge of the first order diamond mode used to determine the pressure is labelled. After laser heating, a high-frequency vibron is observed, associated with free molecular H$_2$ released by thermal decomposition of the surrounding ammonia borane. (b) The derivative of the spectra in (a) in the wavenumber range of the diamond, where the pressure was determined from the wavenumber corresponding to the derivative minima according to the Akahama calibration \cite{akahama_pressure_2006}. (c) The position of the H$_2$ vibron observed in DAC 1 immediately after laser heating. The pressure inferred from the H$_2$ vibron according to Ref. \cite{eremets_universal_2023} is indicated.} 
    \label{fig: DAC1_Raman pressure + vibron}
\end{figure}

\begin{table}[ht!]
\caption{Experimental details for XRD measurements performed on beamlines I15 at Diamond Light Source (DLS), and ID11 and ID27 at the European Synchrotron Radiation Facility (ESRF).}
\centering

\setlength{\tabcolsep}{5pt}
\renewcommand{\arraystretch}{1.2}

\begin{tabular}{|c|c|c|c|c|c|}
\hline

Beamline &
$\lambda$ &
\makecell{Beam diameter\\at FWHM} &
\makecell{Sample-to-detector\\distance} &
Detector &
Calibrant \\

&
(\si{\angstrom}) &
(\si{\micro\metre}) &
(\si{\milli\metre}) &
&
\\

\hline

I15 (DLS) &
0.30996 &
16
& 446
& Pilatus3 X CdTe 2M
& LaB$_6$ \\

ID11 (ESRF) &
0.28436 &
1
& 148
& Eiger2 X CdTe 4M
& CeO$_2$ \\

ID27 (ESRF) &
0.3738 &
0.7
& 200
& Eiger2 X CdTe 9M 
& CeO$_2$ \\

\hline
\end{tabular}

\label{table:xrd_setup}

\end{table}

\end{document}